\documentclass[aps,prb, twocolumn, superscriptaddress]{revtex4-1}

\usepackage{graphicx} 
\usepackage{amssymb} 
\usepackage[usenames]{color}
\usepackage{amsmath}
\usepackage{xspace}
\usepackage{longtable}
\usepackage{soul}
\usepackage{epstopdf}
\usepackage{array}
\usepackage{tabularx}

\begin{document}

\title{Magnetism in a family of $S = 1$ square lattice antiferromagnets Ni$X_2$(pyz)$_2$ ($X = $ Cl, Br, I, NCS; pyz = pyrazine)} 

\author{J. Liu}
\email{junjie.liu@physics.ox.ac.uk}
\affiliation{Department of Physics, Clarendon Laboratory, University of Oxford, Parks Road, Oxford OX1~3PU, UK}

\author{P. A. Goddard}
\affiliation{Department of Physics, University of Warwick, Gibbet Hill Road, Coventry, CV4~7AL, UK}

\author{J. Singleton}
\affiliation{National High Magnetic Field Laboratory, Los Alamos National Laboratory, MS-E536, Los Alamos, NM 87545, USA}

\author{J. Brambleby}
\affiliation{Department of Physics, University of Warwick, Gibbet Hill Road, Coventry, CV4~7AL, UK}

\author{F. Foronda}
\affiliation{Department of Physics, Clarendon Laboratory, University of Oxford, Parks Road, Oxford OX1~3PU, UK}

\author{J. S. M\"{o}ller}
\affiliation{Department of Physics, Clarendon Laboratory, University of Oxford, Parks Road, Oxford OX1~3PU, UK}

\author{Y. Kohama}
\affiliation{National High Magnetic Field Laboratory, Los Alamos National Laboratory, MS-E536, Los Alamos, NM 87545, USA}

\author{S. Ghannadzadeh}
\affiliation{Department of Physics, Clarendon Laboratory, University of Oxford, Parks Road, Oxford OX1~3PU, UK}

\author{A. Ardavan}
\affiliation{Department of Physics, Clarendon Laboratory, University of Oxford, Parks Road, Oxford OX1~3PU, UK}

\author{S. J. Blundell}
\affiliation{Department of Physics, Clarendon Laboratory, University of Oxford, Parks Road, Oxford OX1~3PU, UK}

\author{T. Lancaster}
\affiliation{Centre for Materials Physics, Durham University, South Road, Durham DH1~3LE, UK}

\author{F. Xiao}
\affiliation{Centre for Materials Physics, Durham University, South Road, Durham DH1~3LE, UK}

\author{R. C. Williams}
\affiliation{Centre for Materials Physics, Durham University, South Road, Durham DH1~3LE, UK}

\author{F. L. Pratt}
\affiliation{ISIS Pulsed Muon Facility, STFC Rutherford Appleton Laboratory, Chilton, Didcot, OX11~0QX, UK}

\author{P. J. Baker}
\affiliation{ISIS Pulsed Muon Facility, STFC Rutherford Appleton Laboratory, Chilton, Didcot, OX11~0QX, UK}

\author{K. Wierschem}
\affiliation{School of Physical and Mathematical Sciences, Nanyang Technological University, Singapore 637371, Singapore}

\author{S. H. Lapidus}
\affiliation{X-ray Science Division, Advanced Photon Source, Argonne National Laboratory, Argonne, IL, 60439, USA}

\author{K. H. Stone}
\affiliation{Department of Physics and Astronomy, State University of New York, Stony Brook, NY 11794, USA}

\author{P. W. Stephens}
\affiliation{Department of Physics and Astronomy, State University of New York, Stony Brook, NY 11794, USA}

\author{J. Bendix}
\affiliation{Department of Chemistry, University of Copenhagen, Copenhagen DK-2100, Denmark}

\author{M. R. Lees}
\affiliation{Department of Physics, University of Warwick, Gibbet Hill Road, Coventry, CV4~7AL, UK}

\author{T. J. Woods}
\affiliation{Department of Chemistry and Biochemistry, Eastern Washington University, Cheney, WA 99004, USA}

\author{K. E. Carreiro}
\affiliation{Department of Chemistry and Biochemistry, Eastern Washington University, Cheney, WA 99004, USA}

\author{H. E. Tran}
\affiliation{Department of Chemistry and Biochemistry, Eastern Washington University, Cheney, WA 99004, USA}

\author{C. J. Villa} 
\affiliation{Department of Chemistry and Biochemistry, Eastern Washington University, Cheney, WA 99004, USA}

\author{J. L. Manson}
\email{jmanson@ewu.edu}
\affiliation{Department of Chemistry and Biochemistry, Eastern Washington University, Cheney, WA 99004, USA}

\begin{abstract}
The crystal structures of Ni$X_2$(pyz)$_2$ ($X$ = Cl (\textbf{1}), Br (\textbf{2}), I (\textbf{3}) and NCS (\textbf{4})) were determined at 298~K by synchrotron X-ray powder diffraction. All four compounds consist of two-dimensional (2D) square arrays self-assembled from octahedral NiN$_4$$X_2$ units that are bridged by pyz ligands. The 2D layered motifs displayed by \textbf{1}-\textbf{4} are relevant to bifluoride-bridged [Ni(HF$_2$)(pyz)$_2$]$Z$F$_6$ ($Z$ = P, Sb) which also possess the same 2D layers. In contrast, terminal $X$ ligands occupy axial positions in \textbf{1}-\textbf{4} and cause a staggering of adjacent layers. Long-range antiferromagnetic order occurs below 1.5 (Cl), 1.9 (Br and NCS) and 2.5~K (I) as determined by heat capacity and muon-spin relaxation. The single-ion anisotropy and $g$ factor of \textbf{2}, \textbf{3} and \textbf{4} are measured by electron spin resonance where no zero--field splitting was found. The magnetism of \textbf{1}-\textbf{4} crosses a spectrum from quasi-two-dimensional to three-dimensional antiferromagnetism. An excellent agreement was found between the pulsed-field magnetization, magnetic susceptibility and $T_\textrm{N}$ of \textbf{2} and \textbf{4}. Magnetization curves for \textbf{2} and \textbf{4} calculated by quantum Monte Carlo simulation also show excellent agreement with the pulsed-field data. \textbf{3} is characterized as a three-dimensional antiferromagnet with the interlayer interaction ($J_\perp$) slightly stronger than the interaction within the two-dimensional [Ni(pyz)$_2$]$^{2+}$ square planes ($J_\textrm{pyz}$).
\end{abstract}
\maketitle

\section{Introduction}

Low-dimensional Ni(II) based $S = 1$ antiferromagnets continue to draw much interest from the condensed-matter science community. Since Haldane\cite{HaldanePL1983,HaldanePRL1983} predicted that an antiferromagnetic Heisenberg chain has a singlet ground state and a finite gap to the lowest excited state for integer spins, this conjecture has inspired numerous studies of $S = 1$ antiferromagnets in low-dimensions. While most of the work done so far is related to one-dimensional (1D) models or quasi-one-dimensional (Q1D) compounds\cite{RenardEPL1987,AjiroPRL1989,AffleckPRB1991,ZapfPRL2006,ZvyaginPRL2007,KohamaPRL2011,OrendacPRB1995,OrendacPRB1999,AlbuquerquePRB2009}, less work has been performed on two-dimensional models (2D) or quasi-two-dimensional (Q2D) compounds\cite{BishopJPhysCondMatt2008,HamerPRB2010,WierschemPRB2012,ZhangPRB2013} partially due to the difficulty of applying theoretical/numerical techniques to these models. In low-dimensional $S = 1$ antiferromagnets, the nature of the ground state can be strongly modified by the spatial dimensionality as well as the zero-field splitting (ZFS) of Ni(II),\cite{WierschemPRL2014} both of which can be tuned by chemical synthesis. In addition, the presence of two orthogonal magnetic orbitals in octahedral coordinated Ni(II), $d_{z^2}$ and $d_{x^2-y^2}$, affords multiple options for forming spin exchange pathways, allowing flexibility in tuning the magnetic dimensionality via crystal engineering. 

We and others have been developing two-dimensional Cu(II)-based square lattices comprised of pyrazine (pyz) bridges. Among these are [Cu(HF$_2$)(pyz)$_2$]$Z$ ($Z$ = BF$_4^-$, PF$_6^-$, SbF$_6^-$ and TaF$_6^-$),\cite{MansonChemComm2006,CizmarPRB2010,MansonJACS2009,MansonJLowTempPhys2010} Cu(ClO$_4$)$_2$(pyz)$_2$,\cite{ChoiChemMat2003,LancasterPRB2007} Cu(BF$_4$)$_2$(pyz)$_2$,\cite{WoodwardInorgChem2007} and [Cu(pyz)$_2$(pyO)$_2$](PF$_6$)$_2$\cite{GoddardPRL2012} which all display long-range order (LRO) between 1.5 and 4.3~K. The square [Cu(pyz)$_2$]$^{2+}$ planes in [Cu(HF$_2$)(pyz)$_2$]$Z$ are connected by HF$_2^-$ bridges to afford three-dimensional (3D) frameworks with $Z$ occupying the interior sites. However the magnetism is very two-dimensional as a result of very weak couplings through Cu-FHF-Cu bonds\cite{GoddardNewJPhys2008} due to limited overlap between the fluorine $p_z$ orbital and the magnetic orbital of Cu(II), $d_{x^2-y^2}$, lying in the [Cu(pyz)$_2$]$^{2+}$ planes.\cite{SteelePRB2011} The last three examples above contain axial ClO$_4^-$, BF$_4^-$ or pyO ligands and the 2D layers stack in a staggered fashion. Extension of some of this work to include Ni(II) has proven to be more challenging as growth of single crystals is difficult. As such, implementation of synchrotron X--ray diffraction to determine crystal structures, including those described here, has been crucial to our characterization efforts. In addition, the $^3$A$_{\textrm{2g}}$ ground state of an octahedrally coordinated Ni(II) ion is magnetically more complex than Cu(II) owing to the presence of ZFS induced by spin-orbital couplings. The effective spin Hamiltonian ($S = 1$) is given by:
\begin{equation}
\label{HwithD}
\hat{H} = \sum\limits_{\left<i\right>} D\hat{S}_{i}^{z^2} + \sum\limits_{\left<i,j\right>} J_{ij}\hat{\mathbf{S}}_i\cdot\hat{\mathbf{S}}_j.
\end{equation}
Experimentally, it becomes difficult to distinguish between the effects from magnetic exchange interactions ($J_{ij}$) and single-ion ZFS ($D$), especially when polycrystalline samples are involved.\cite{MansonInorgChem2011} The difficulty lies in that in many circumstances magnetometry data can be fitted to several models with different combinations of $D$ and $J$, which makes it challenging to characterize a system unambiguously. In which case, additional spectroscopic measurements are required to constrain the parameters in the Hamiltonian.

Considering these challenges, we recently described the structural, electronic and magnetic properties of [Ni(HF$_2$)(pyz)$_2$]$Z$ ($Z$ = PF$_6^-$, SbF$_6^-$)\cite{MansonInorgChem2011,MansonDaltTran2012}. Interestingly, $Z$ = PF$_6^-$ exists as two isolable polymorphs with similar 3D structural motifs; the $\alpha$-phase is monoclinic while the $\beta$-phase is tetragonal and isostructural to the equivalent Cu(II) compound. A spatial exchange anisotropy was found in these materials due to the presence of co-existing Ni-FHF-Ni ($J_{\textrm{FHF}}$) and Ni-pyz-Ni pathways ($J_{\textrm{pyz}}$), where $J_{\textrm{FHF}} > J_{\textrm{pyz}}$. The dominant Ni-FHF-Ni pathways allowed us to interpret the $\chi(T)$ data according to a Q1D chain model above $T_{\textrm{max}}$ but it was not possible to experimentally determine $J_{\textrm{pyz}}$ owing to the polycrystalline nature of the samples. Density-functional theory (DFT) confirmed the magnetic exchange properties of these systems and that $J_{\textrm{pyz}}$ was indeed much smaller than $J_{\textrm{FHF}}$. Angular Overlap Model (AOM) analyses of UV-Vis spectroscopic data determined $D$ to be -7.5~K ($\alpha$-PF$_6^-$), 10.3~K ($\beta$-PF$_6^-$) and 11.2~K (SbF$_6^-$).\cite{MansonInorgChem2011} The correspondingly high $T_{\textrm{N}}$ of 6.2, 7.0 and 12.2~K suggest that $J_{\textrm{pyz}}$ must be larger than that calculated or, alternatively, the magnetic orders are assisted by $D$. In order to address these scenarios as well as find $J_{\textrm{pyz}}$ quantitatively, analogous model compounds based on weakly interacting 2D [Ni(pyz)$_2$]$^{2+}$ square lattices are required for comparison. 

Four compounds with similar [Ni(pyz)$_2$]$^{2+}$ square lattices have been synthesized and studied:\\*
\indent \quad \textbf{1} NiCl$_2$(pyz)$_2$\\
\indent \quad \textbf{2} NiBr$_2$(pyz)$_2$\\
\indent \quad \textbf{3} NiI$_2$(pyz)$_2$\\
\indent \quad \textbf{4} Ni(NCS)$_2$(pyz)$_2$\\*
The simple compounds \textbf{1}, \textbf{2} and \textbf{4} were synthesized and spectroscopically characterized many years ago\cite{LeverJChemSoc1963,LeverJChemSoc1964,GoldsteinJCHemSocDaltTran1972,OtienoCanJChem1995} although their crystal structures were not determined explicitly. More recently, the structure of \textbf{2} was determined by powder neutron diffraction and found to be consistent with the hypothetical square lattice structure.\cite{JamesAustJChem2002} A related Ni(II) compound, \textbf{4}, reportedly exists in two polymorphic forms, however, as will be described below, we find only one of the two structures present in our samples.\cite{WriedtEJIC2009,WangZAAC2010}

As for the magnetic properties of \textbf{1}-\textbf{4}, the temperature dependence of the magnetic susceptibility data, $\chi(T)$, for \textbf{1} and \textbf{2} have been reported ($T \ge 5~\textrm{K}$)\cite{OtienoCanJChem1995,JamesAustJChem2002} while those for \textbf{3} and \textbf{4} have not. The analysis of the $\chi(T)$ data for \textbf{1} and \textbf{2} gave $D = 7.92$ and 14.8~K, respectively. Furthermore, these studies also suggested that magnetic couplings along Ni-pyz-Ni were probably very weak. An estimate of $J_\textrm{pyz}$ was made by employing a mean-field contribution, giving $zJ$ = 0.39~K for \textbf{1} and 0.95~K for \textbf{2}.\cite{OtienoCanJChem1995,JamesAustJChem2002} Compound \textbf{3} has not been reported previously and we describe it here for the first time.

In this work, we have carried out an extensive experimental and theoretical investigation of \textbf{1}-\textbf{4}, employing modern instrumental methods to characterize their structural as well as temperature and field-dependent magnetic properties. Our interpretation of the experimental results suggests the interlayer magnetic couplings in \textbf{1}-\textbf{4} are significantly suppressed compared to the [Ni(HF$_2$)(pyz)$_2$]$Z$ compounds and become comparable or less than $J_{\textrm{pyz}}$. To clarify the possible Ni(II) ZFS contribution to the magnetism, electron spin resonance measurements were performed on \textbf{1}-\textbf{4}. $J_{\textrm{pyz}}$ in \textbf{2}-\textbf{4} is quantitatively determined within the picture of Q2D magnetism and the conclusions are supported by quantum Monte Carlo (QMC) calculations. The common [Ni(pyz)$_2$]$^{2+}$ square lattices exhibited by \textbf{1}-\textbf{4} are relevant to establishing magnetostructural correlations in the metal-organic frameworks, [Ni(HF$_2$)(pyz)$_2$]$Z$ ($Z$ = PF$_6^-$, SbF$_6^-$).

\section{Experimental methods}

\textit{Syntheses}. Following a general procedure, \textbf{1} and \textbf{2} were prepared as powders using a fast precipitation reaction between the corresponding NiX$_2\cdot$4H$_2$O and two equivalents of pyrazine. Each reagent was dissolved in 3~mL of H$_2$O and quickly mixed together while stirring. For \textbf{4}, KNCS (2.16~mmol, 0.2100~g) and pyz (2.16~mmol, 0.1730~g) were dissolved together in 5~mL of H$_2$O. To this solution was added, while stirring, Ni(NO$_3$)$_2$$\cdot$yH$_2$O (1.08~mmol, 0.1973~g) to afford a pale blue precipitate. In all instances, the powders were isolated by suction filtration, washed with H$_2$O, and dried in vacuo for $\sim$2 hours. Compound \textbf{3} was prepared via a mechanochemical reaction involving grinding of NiI$_2$ (2.88~mmol, 0.9013~g) with an excess of pyrazine (6.78~mmol, 0.2307~g). A Parr acid-digestion bomb was charged with the reaction mixture and placed inside a temperature programmable oven which was set at a temperature of 403 K. The sample was held isothermal for 2 weeks and then allowed to cool slowly to room temperature at which time a homogeneous orange-brown solid had formed. The final product was obtained by washing the sample with fresh diethyl ether to remove any unreacted pyz. All four compounds were highly pure and isolated in yields exceeding 90\%. 

\begin{table*}[t]
\caption{Crystallographic refinement parameters for \textbf{1}-\textbf{4} as determined by synchrotron X-ray powder diffraction.}
\centering
\begin{tabular}{c c c c c}
\hline
Compound & NiCl$_2$(pyz)$_2$ (\textbf{1}) & NiBr$_2$(pyz)$_2$ (\textbf{2}) & NiI$_2$(pyz)$_2$ (\textbf{3}) & Ni(NCS)$_2$(pyz)$_2$ (\textbf{4}) \\
\hline
Emp. Formula & C$_8$N$_4$H$_8$NiCl$_2$ & C$_8$N$_4$H$_8$NiBr$_2$ &  C$_8$N$_4$H$_8$NiI$_2$ & C$_{10}$N$_6$H$_8$NiS$_2$ \\
Wt. (g/mol) & 289.77 & 378.67 & 472.68 & 335.03 \\
$T$ (K) & 298 & 298 & 100 & 298 \\
Crystal Class & tetragonal & tetragonal & tetragonal & monoclinic \\
Space group & $I4/mmm$ & $I4/mmm$ & $I4/mmm$ & $C2/m$ \\
$a$ (\AA) & 7.0425(2) & 7.0598(2) & 7.057502(18) & 9.9266(2) \\
$b$ (\AA) & 7.0425(2)	 & 7.0598(2) & 7.057502(18) & 10.2181(2) \\
$c$ (\AA) & 10.7407(3) & 11.3117(3) & 12.25594(5) & 7.2277(2) \\
$\beta$ ($^\circ$) &90 & 90 & 90 & 118.623(2) \\
$V$ (\AA$^3$) & 532.71(3) & 563.79(4) & 610.448(5) & 643.52(3) \\
$Z$ & 2 & 2 & 2 & 2\\
$\rho$ (g/cm$^3$) & 1.807 & 2.231 & 2.571 & 1.729 \\
$\lambda$ (\AA) & 0.699973 & 0.754056 & 0.41374 & 0.6984 \\
$R_{\textrm{WP}}$ & 0.05592 & 0.04524 & 0.04648 & 0.04531 \\
$R_{\textrm{exp}}$ & 0.06987 & 0.05449 & 0.03249 & 0.05644 \\
$\chi$ & 1.471 & 1.093 & 1.431 & 1.720 \\
\hline
\end{tabular} 
\label{TableStruc1}
\end{table*}

\textit{Structural determinations}. For Ni$X_2$(pyz)$_2$ ($X$ = Cl, Br or NCS), high resolution synchrotron powder X-ray diffraction patterns were collected at the X12A and X16C beamline at the National Synchrotron Light Source at Brookhaven National Laboratory. X-rays of a particular wavelength were selected using a Si(111) channel cut monochromator. Behind the sample, the diffracted beam was analyzed with a Ge(111) crystal and detected by a NaI scintillation counter. Wavelength and diffractometer zero were calibrated using a sample of NIST Standard Reference Material 1976, a sintered plate of Al$_2$O$_3$.  The sample was loaded into a 1.0~mm diameter glass capillary and flame sealed. 

For NiI$_2$(pyz)$_2$, high resolution synchrotron powder X-ray diffraction data were collected using beamline 11-BM at the Advanced Photon Source (APS).\cite{WangRevSciInstr2008} Discrete detectors are scanned over a 34$^\circ$ in $2\theta$ range, with data points collected every $0.001^\circ$ and the scan speed of $0.01^\circ$/s. Data are collected while continually scanning the diffractometer $2\theta$ arm.

Indexing was performed in TOPAS Academic\cite{TOPAS,CoelhoJApplCrystal2000} and space groups were tentatively assigned through systematic absences to be $I4/mmm$ for Ni$X_2$(pyz)$_2$ ($X$ = Cl, Br or I) and $C2/m$ for Ni(NCS)$_2$(pyz)$_2$. From the space group assignment and stoichiometric contents, it is possible to place the Ni on a corresponding special position. The rest of the atomic positions can be determined through simulated annealing in TOPAS Academic. From these initial models, these structures were successfully refined to determine more precise atomic positions. Pyrazine hydrogens were placed on ideal geometrically determined positions.

\textit{Magnetic measurements}. Magnetization ($M$) versus temperature data were collected (and converted to susceptibility by the relation $\chi(T) = M/H$) on a Quantum Design MPMS 7~T SQUID. Powder samples of \textbf{1}-\textbf{4} were loaded into gelatin capsules, mounted in a plastic drinking straw, and affixed to the end of a stainless steel/brass rod. The sample was cooled in zero-field to a base temperature of 2~K, the magnet charged to 0.1~T, and data taken upon warming to 300~K. All data were corrected for core diamagnetism using tabulated data. 

\textit{Pulsed-fields} $M(B)$ measurements (up to 60~T) made use of a 1.5~mm bore, 1.5~mm long, 1500-turn compensated coil susceptometer, constructed from a 50 gauge high-purity copper wire. When the sample is within the coil, the signal voltage $V$ is proportional to $\mathrm{d}M/\mathrm{d}t$, where $t$ is the time. Numerical integration of $V$ is used to evaluate $M$. The sample is mounted within a 1.3~mm diameter ampule that can be moved in and out of the coil. Accurate values of $M$ are obtained by subtracting empty coil data from that measured under identical conditions with the sample present. The susceptometer was placed inside a $^3$He cryostat providing a base temperature of 0.5~K. The field $B$ was measured by integrating the voltage induced in a 10-turn coil calibrated by observing the de Haas-van Alphen oscillations of the belly orbits of the copper coils of the susceptometer. 

\textit{Heat capacity}. $C_\textrm{p}$ measurements were carried out on polycrystalline samples of \textbf{1}-\textbf{4} by means of two independent techniques; the traditional relaxation\cite{BachmannRevSciInstr1972} and dual-slope methods\cite{RiegelJPhysE1986}. In the relaxation method, the heat pulse was applied to the sample heater, and the resultant exponentional temperature decay with a small temperature step, which is $\sim3\%$ of the thermal bath temperature, was observed. The $C_\textrm{p}$ at a single temperature was evaluated by the time constant of the decay curve and the thermal conductance of the thermal link. In the dual slope method, the sample was heated and subsequently cooled through a broad temperature range, and the $C_\textrm{p}(T)$ in the wide temperature range was evaluated using both heating and cooling curves. This method allows quick collection of a large amount of $C_\textrm{p}(T)$ data, which is important in determining the transition temperature ($T_\textrm{N}$) at several magnetic fields. However, it requires an excellent thermal contact between the sample and the thermometer, that can only be used in cases of minimal tau-2 effects, i.e. the thermal relaxation between the sample and the platform must be fast\cite{RiegelJPhysE1986}. For this reason,  $C_\textrm{p}(T)$ of \textbf{1} was obtained by traditional relaxation method only. For \textbf{4}, using the same set-up as the  $C_\textrm{p}$ experiments, we additionally observed a magnetocaloric effect (MCE) by sweeping the magnetic field at 1~T/min. This method measures the entropy change as a function of magnetic field and can detect phase boundaries with cooling and heating responses.\cite{Tishin2003} These $C_\textrm{p}(T)$ and magnetocaloric effect (MCE) measurements were performed on 2.910, 1.479, 2.284 and 0.3406~mg of \textbf{1}, \textbf{2}, \textbf{3} and \textbf{4}, respectively. The powders were mixed with a small amount of Apiezon-N grease and pressed between Si plates to obtain good temperature homogeneity. \textbf{1}, \textbf{2} and \textbf{4} were measured in an Oxford 15~T superconducting magnet system capable of reaching a base temperature of 0.4~K. \textbf{3} was measured in a 9~T Quantum Design Physical Property Measurement System. The addenda specific heat due to Apiezon-N grease, Si plates, and sample platform were measured separately. After subtracting the addenda contribution from the total specific heat, the specific heat of the sample was obtained. Excellent agreement (within $\sim 5 \%$) between the two $C_\mathrm{p}(T)$ techniques was confirmed for \textbf{2} and \textbf{4}. 

\textit{Muon-spin relaxation}. Zero-field muon-spin relaxation (ZF $\mu$SR) measurements were made on a polycrystalline samples of {\bf 1}-{\bf 4} using the General Purpose Surface (GPS) spectrometer at the Swiss Muon Source ({\bf 1} and {\bf 2}), and the EMU ({\bf 1}), MuSR ({\bf 3}) and ARGUS ({\bf 4}) instruments at the STFC ISIS facility. For the measurement the samples were mounted in silver foil packets onto silver backing plates.  

In a $\mu$SR experiment \cite{BlundellContempPhys1999} spin-polarized positive muons are stopped in a target sample, where the muon usually occupies an interstitial position in the crystal. The observed property in the experiment is the time evolution of the muon spin polarization, the behavior of which depends on the local magnetic field at
the muon site. Each muon decays, with an average  lifetime of 2.2~$\mu$s, into two neutrinos and a positron, the latter particle being emitted preferentially along the instantaneous direction of the muon spin. Recording the time dependence of the positron emission directions therefore allows the determination of the spin-polarization of the ensemble of muons. In our experiments positrons are detected by detectors placed forward (F) and backward (B) of the initial muon polarization direction. Histograms $N_{\mathrm{F}}(t)$ and $N_{\mathrm{B}}(t)$ record the number of positrons detected in the two detectors as a function of time following the muon implantation. The quantity of interest is the decay positron asymmetry function, defined as
\begin{equation}
A(t)=\frac{N_{\mathrm{F}}(t)-\alpha_{\mathrm{exp}} N_{\mathrm{B}}(t)}
{N_{\mathrm{F}}(t)+\alpha_{\mathrm{exp}} N_{\mathrm{B}}(t)} \, ,
\end{equation}
where $\alpha_{\mathrm{exp}}$ is an experimental calibration constant. $A(t)$ is proportional to the spin polarization of the muon ensemble. 

\textit{Electron spin resonance} (ESR). D-band (130~GHz) ESR measurements were performed on powder samples of \textbf{1}-\textbf{3}. A phase-locked dielectric resonator oscillator in conjunction with a series of IMPATT diodes were used as the microwave source and detector. A field modulation was employed for D-band ESR measurements. Multi-high-frequency EPR measurements were also performed on a powder sample of \textbf{2}-\textbf{4} using a cavity perturbation technique spanning the frequency range from 40 to 170~GHz. A millimeter-vector-network-analyzer served as the microwave source and detector. ESR measurements were performed in a 6~T horizontal-bore superconducting magnet with the temperature regulated between 1.5~K and 300~K using a helium gas flow cryostat.

\textit{Quantum Monte Carlo calculations}. Numerical calculations of the spin-1 antiferromagnetic Heisenberg model in an applied magnetic field were performed using the stochastic series expansion quantum Monte Carlo (QMC) method with directed loop updates.\cite{SyljuasenPRE2002} For antiferromagnetic exchange interactions, sublattice rotation is required to avoid the sign problem in QMC. By taking the direction of the applied magnetic field as the discretization axis, sublattice rotation on a bipartite lattice leads to a sign problem free Hamiltonian as long as the applied field is parallel or perpendicular to the axis of exchange anisotropy. The case of applied field parallel to the axis of exchange anisotropy has been well-studied. For the case of perpendicular applied fields, we use a slightly modified approach to account for a lack of the usual conservation law.\cite{SyljuasenPRE2003}

\textit{Density Functional Theory} (DFT). Computational modeling was performed on dinuclear entities using the structural data from X-ray determinations. Evaluation of the exchange couplings was based on the broken-symmetry (BS) approach of Noodleman\cite{NoodlemanJChemPhys1981} as implemented in the ORCA ver.2.8 suite of programs.\cite{NeeseORCA,SinneckerJBiolChem2005,NeeseCoordChemRev2009} The formalism of Yamaguchi, which employs calculated expectation values $\langle S^2\rangle$ for both high-spin and broken-symmetry states, was used.\cite{Yamaguchi1986,SodaChemPhysLett2000} Calculations related to magnetic interactions have been performed using the PBE0 functional. The def2-TZVP basis function set from Ahlrichs was used.\cite{WeigendPCCP2005} 

\begin{figure}[t]
\centering
\includegraphics[width=\columnwidth]{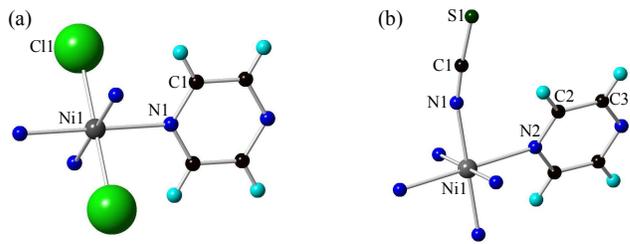}
\caption{Room temperature asymmetric units and atom labeling schemes for (a) NiCl$_2$(pyz)$_2$ (\textbf{1}) and (b) Ni(NCS)$_2$(pyz)$_2$ (\textbf{4}). The asymmetric units and atom labeling schemes for NiBr$_2$(pyz)$_2$ (\textbf{2}) and NiI$_2$(pyz)$_2$ (\textbf{3}) are similar to those of \textbf{1} with the Cl atom being replaced by Br and I for \textbf{2} and \textbf{3}, respectively.}
\label{figlabel}
\end{figure}

\section{Results}
\subsection{Crystal structures}

Crystallographic refinement details as well as selected bond lengths and bond angles for \textbf{1}-\textbf{4} are listed in Tables \ref{TableStruc1} and \ref{TableStruc2}. The data correspond to room temperature (\textbf{1}, \textbf{2}, and \textbf{4}) and 100 K (\textbf{3}) structures.

\begin{figure}[t]
\centering
\includegraphics[width=\columnwidth]{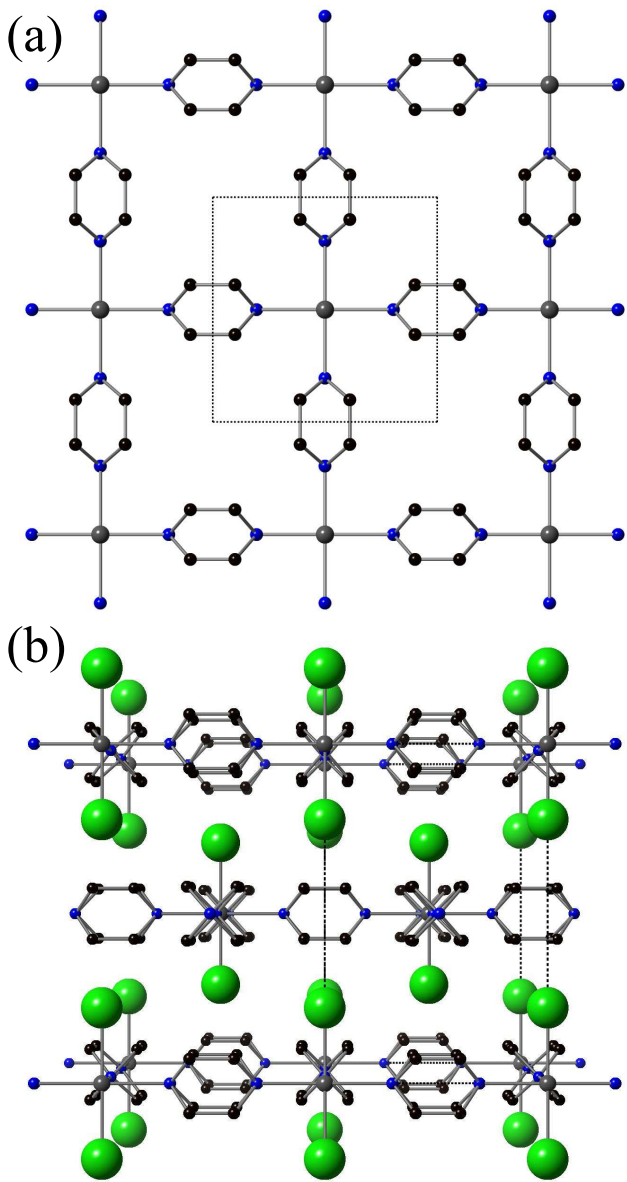}
\caption{(a) Two-dimensional layer of  NiCl$_2$(pyz)$_2$ (\textbf{1}) with axial Cl atoms omitted for clarity. (b) Staggered packing of 2D layers in \textbf{1}. The positional disorder of the pyz ligands is shown as the double pyz rings connecting Ni atoms. NiBr$_2$(pyz)$_2$ (\textbf{2}) and NiI$_2$(pyz)$_2$ (\textbf{3}) are isostructural with \textbf{1}. The unit cell is indicated by dashed lines. Ni, Cl, N and C atoms are represented as gray, green, blue and black spheres, respectively. H atoms are omitted for clarity.}
\label{figXray1}
\end{figure}

\textit{NiCl$_2$(pyz)$_2$ } (\textbf{1}), \textit{NiBr$_2$(pyz)$_2$} (\textbf{2}) and \textit{NiI$_2$(pyz)$_2$} (\textbf{3}). The atom labeling scheme is shown in Fig.\,\ref{figlabel}(a). \textbf{1}-\textbf{3} are isomorphous and consist of tetragonally-elongated Ni$X_2$N$_4$ sites, with the axial sites being occupied by the bulkier $X$ anions. The Ni1-N distances are only slightly perturbed by $X$ [2.145(2)~\AA\ (\textbf{1}), 2.131(4)~\AA\ (\textbf{2}) and 2.133(1)~\AA\ (\textbf{3})] whereas the Ni1-$X$ bond lengths are substantially longer at 2.400(1)~\AA\ (\textbf{1}), 2.5627(9)~\AA\ (\textbf{2}) and 2.7919(1)~\AA\ (\textbf{3}) due to increasing ionic radius of the halide. The Ni-N distances in \textbf{1}-\textbf{3} are similar to those reported in other compounds with Ni-pyz-Ni bridges. However, the axial bonds (Ni-\textit{X}) in \textbf{1}-\textbf{3} are significantly longer than those in compounds with related structures that contain either 1D or 2D Ni-pyz-Ni bridges. This is likely due to the relatively large radius of the halogen atoms in \textbf{1}-\textbf{3} comparing the axial ligands in other systems which contain smaller O or N-donor atoms. The topological structures of \textbf{1}-\textbf{3} can be described as infinite 2D square lattices with Ni$X_2$N$_4$ octahedra bridged by pyz linkages along the $a$- and $b$-axes [Fig.\,\ref{figXray1}(a)] to afford perfectly linear Ni-N$\cdots$N trajectories. The four-fold rotational symmetry about the $c$-axis as well as the mirror planes of the $I4/mmm$ space group lead to two-fold positional disorder of the pyz ligands that surround the Ni ion [Fig.\,\ref{figXray1}(b)]. The pyz ligands have equal probability of appearing in one of the two positions. Similar pyz disorder has been found in Ni(OCN)$_2$(pyz)$_2$ which possesses the same $I4/mmm$ space group.\cite{WangZAAC2010} The canting angle at which the pyz rings are tilted about their N-N axes with respect to the $ab$-plane are essentially the same (47.4$^\circ$, 46.5$^\circ$ and 45.8$^\circ$ for \textbf{1}, \textbf{2} and \textbf{3}, respectively); by contrast, these values are significantly different to that found for \textbf{4} (65.3$^\circ$).

The [Ni(pyz)$_2$]$^{2+}$ layers stack along the $c$-direction such that the Ni(II) ion of a given lattice lies above/below the centers of neighboring square lattices [Fig.\,\ref{figXray1}(b)]. The bulky $X$ anions act as spacers to separate each layer, giving interlayer Ni$\cdots$Ni separations of 7.32~\AA\ (\textbf{1}), 7.54~\AA\ (\textbf{2}) and 7.90~\AA\ (\textbf{3}). It should be noted that the 2D structural motif was anticipated based on early infrared spectroscopic evidence\cite{GoldsteinJCHemSocDaltTran1972,OtienoCanJChem1995} and now confirmed here using structural data. 

Two structures were reported for the Co-congener of \textbf{1}, CoCl$_2$(pyz)$_2$. The far-infrared spectra for CoCl$_2$(pyz)$_2$ suggested tetragonal symmetry ($I4/mmm$)\mbox{\cite{CarreckJchemCommD1971}} whereas later X-ray study revealed an orthorhombic space group \textit{Ccca}\mbox{\cite{GairingZKristallogr1996}}. Both structures consist of a square lattice motif with Co(II) centers bridged by pyz ligands. For the sake of comparison, the synchrotron diffraction data for \textbf{1} and \textbf{2} were also fitted with the \textit{Ccca} space group but no good agreement was found for this orthorhombic space group.

\begin{figure}[b]
\centering
\includegraphics[width=\columnwidth]{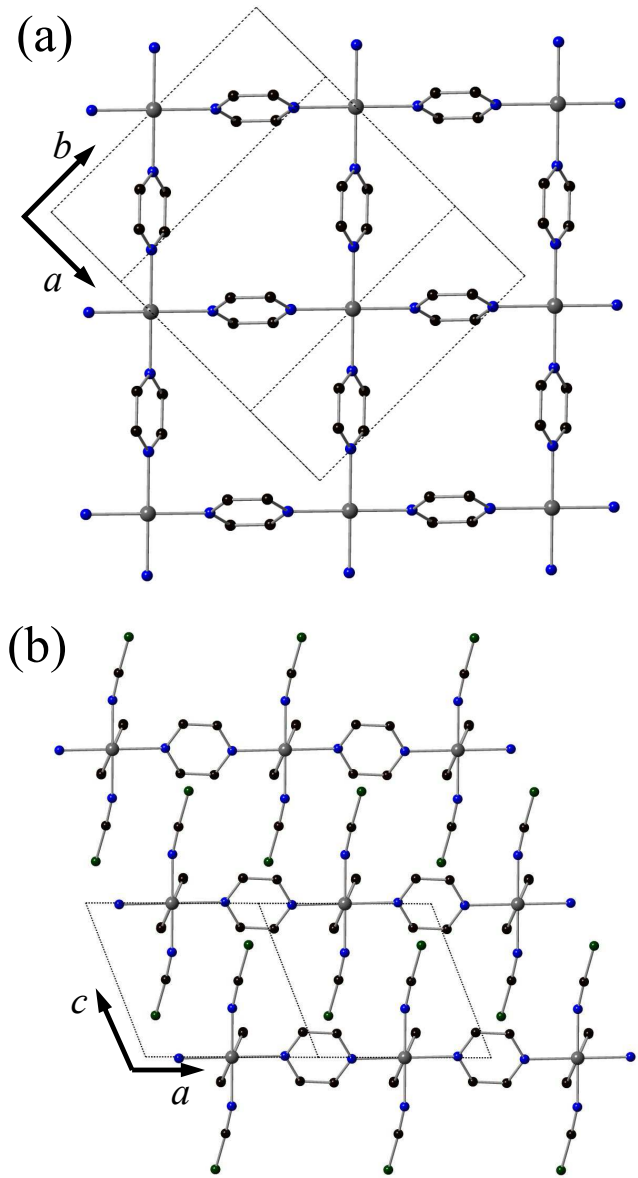}
\caption{Crystal structure of Ni(NCS)$_2$(pyz)$_2$ (\textbf{4}). (a) A 2D sheet viewed normal to the $ab$-plane where the slight rhombic distortion of the sheet is readily seen. NCS ligands are omitted for clarity. (b) Staggered packing of sheets. The unit cell is indicated by dashed lines. Ni, S, N and C atoms are represented as gray, dark green, blue and black spheres, respectively. H atoms are omitted for clarity.}
\label{figXray2}
\end{figure}

\textit{Ni(NCS)$_2$(pyz)$_2$} (\textbf{4}). Previously, two different structural modifications have been reported,\cite{WriedtEJIC2009,WangZAAC2010} each having monoclinic symmetry ($C2/m$ and $P21/n$) at 293~K. Although both structures possess octahedral Ni(II) centers, four pyz ligands in the equatorial plane, two axial NCS$^-$ ligands and 2D layered motifs that consist of orthogonally cross-linked Ni-pyz-Ni chains, an essential difference between them lies in the relative distortion of the NiN$_6$ octahedron. In the $C2/m$ structure as described by Wriedt \textit{et al.},\cite{WriedtEJIC2009} four equivalent Ni-N$_{\textrm{pyz}}$ bonds [2.162(1) \AA] occupy the 2D plane while the axial direction contains shorter Ni-N bonds [2.033(2) \AA]. In contrast, three distinct pairs of Ni-N distances are found in the $P21/n$ variant, with an axial elongation along one of the Ni-pyz-Ni chains [Ni-N$_{\textrm{pyz}}$ = 2.440(3) \AA]. The other two Ni-N bonded pairs contain the other (orthogonal) Ni-pyz-Ni chain whereas the Ni-N bonds (from the NCS$^-$ ligand) are 1.945(3) \AA.  

\begin{table}[t]
\caption{Selected bond lengths (\AA) and bond angles ($^\circ$) for \textbf{1}-\textbf{4}.}
\centering
\begin{tabular}{l >{\hfill}p{2cm} || l r}
\hline
\multicolumn{4}{c}{NiCl$_2$(pyz)$_2$ (\textbf{1})}\\
\hline
Ni1-N1 & 2.145(2) & Ni1-Cl1 & 2.400(1) \\
N1-C1 & 1.336(2) & N1-Ni1-Cl1 & 90$^\circ$ \\
Cl1-Ni1-Cl1 & 180$^\circ$ & N1-Ni1-N1 & 90$^\circ$ \\
Ni1-N1-C1 & 120.5(1)$^\circ$ & Dihedral angle$^a$ & 47.4(2)$^\circ$ \\
\hline
\multicolumn{4}{c}{NiBr$_2$(pyz)$_2$ (\textbf{2})}\\
\hline
Ni1-N1 & 2.131(4) & Ni1-Br1 & 2.5627(9) \\
N1-C1 & 1.351(3) & N1-Ni1-Br1 & 90$^\circ$ \\
Br1-Ni1-Br1 & 180$^\circ$ & N1-Ni1-N1 & 90$^\circ$ \\
Ni1-N1-C1 & 121.4(2)$^\circ$ & Dihedral angle$^a$ & 46.5(2)$^\circ$ \\
\hline
\multicolumn{4}{c}{NiI$_2$(pyz)$_2$ (\textbf{3})}\\
\hline
Ni1-N1 & 2.133(1) & Ni1-I1 & 2.7919(1) \\
N1-C1 & 1.349(1) & N1-Ni1-I1 & 90$^\circ$ \\
I1-Ni1-I1 & 180$^\circ$ & N1-Ni1-N1 & 90$^\circ$ \\
Ni1-N1-C1 & 121.4(2)$^\circ$ & Dihedral angle$^a$ & 45.8(1)$^\circ$ \\
\hline
\multicolumn{4}{c}{Ni(NCS)$_2$(pyz)$_2$ (\textbf{4})}\\
\hline
Ni1-N1 & 2.020(5) & Ni1-N2 & 2.184(3) \\
N1-C1 & 1.184(7) & N2-C2 & 1.303(3) \\
S1-C1 & 1.591(5) & C2-C3 & 1.401(5) \\
N1-C2-S1 & 175.5(7)$^\circ$ & Ni1-N1-C1 & 163.3(5)$^\circ$ \\
N1-Ni1-N2 & 88.4(2)$^\circ$ & N1-Ni1-N1 & 180$^\circ$ \\
N2-Ni1-N2 & 180$^\circ$ &Dihedral angle$^a$ & 65.3(2)$^\circ$ \\
\hline
\multicolumn{4}{l}{$^a$ Measured as the pyz tilt angle relative to the $ab$-plane.}\\
\end{tabular}
\label{TableStruc2}
\end{table}

For the sake of a careful structural and magnetic comparison to \textbf{1}-\textbf{3} we have re-examined the 298~K structure of \textbf{4} using high-resolution synchrotron powder X-ray diffraction. We found the crystal structure of \textbf{4} to be essentially identical to that of the reported $C2/m$ phase and describe the structure in detail here as it is pertinent to the development of magnetostructural correlations.

Indeed, \textbf{4} features four equivalent Ni-N2 (from pyz) bond distances of 2.184(3)~\AA\ while Ni-N1 (from NCS$^-$) are shorter at 2.020(5) \AA. These Ni-N distances are significantly different to the $P21/n$ phase. Other striking variations are observed in the bond angles about the NiN$_6$ octahedron. The main structural feature of \textbf{4} is the planar 2D nearly square grid that propagates in the $ab$-plane as illustrated in Fig.\,\ref{figXray2}(a). Here, adjoining orthogonal chains afford equivalent intralayer Ni$\cdots$Ni separations of 7.123(1)~\AA\ along both Ni-pyz-Ni chains. The square exhibits a slight rhombic distortion such that the diagonals vary by 3\% (9.926 vs 10.218~\AA). Also of importance is that the pyz ligands form slightly nonlinear Ni-pyz-Ni bridges such that the N-donor atoms (N1) of the pyz ring lie just off the Ni$\cdots$Ni trajectory. The Ni1-N2$\cdots$Ni1 backbone has an angle of 177.3$^\circ$ as compared to the 180$^\circ$ angles found in \textbf{1}-\textbf{3}. By comparison, the $P21/n$ structure exhibits inequivalent Ni$\cdots$Ni distances of 6.982(1)~\AA\ along the $a$-axis and 7.668(2)~\AA\ along $b$. 

The 2D layers in \textbf{4} are staggered such that the axial NCS$^-$ ligands protrude toward the midpoints of adjacent layers; they stack perpendicular to the $c$-axis [Fig.\,\ref{figXray2}(b)]. The closest interlayer Ni$\cdots$Ni separation is 7.2277(2)~\AA\ which corresponds to the $c$-axis repeat unit. 

An isomorphous series of compounds exists, $M$(NCS)$_2$(pyz)$_2$ where $M$ = Mn, Fe, Co, and Ni.\cite{RealInorgChem1991,LuInorgChem1997,LloretMCLC1999} Cu(II) ion forms Cu(NCS)$_2$(pyz) which contains 2D rectangular layers made up of bi-bridged Cu-(NCS)$_2$-Cu ribbons that are cross-linked via pyz bridges.\cite{BordalloPolyhedron2003} Substitution of 4,4'-bipyridine (4,4'-bipy) for pyz affords the related structure Cu(NCS)$_2$(4,4'-bipy).\cite{LuoActCrySecE2006}

\subsection{Search for long range ordering with heat capacity}
\label{SectionCp}

\begin{figure}[t]
\centering
\includegraphics[width=\columnwidth]{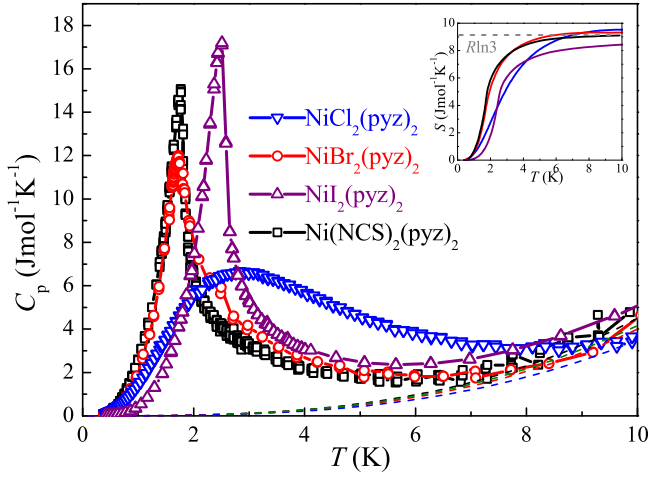}
\caption{Specific heat of polycrystalline samples of Ni$X_2$(pyz)$_2$ ($X$ = Cl (\textbf{1}), Br (\textbf{2}), I (\textbf{3}) and NCS (\textbf{4})). Main panel: zero field heat capacity data collected between 1-10~K. The dash lines represent the estimated lattice contribution $C_{\textrm{latt}}$. Inset: the temperature dependence of the magnetic entropy for \textbf{1}-\textbf{4}.}
\label{figCP1}
\end{figure}

Fig.~\ref{figCP1} displays the zero-field heat capacity ($C_{\textrm p}$) of compounds {\textbf 1}-{\textbf 4} collected in the temperature range of 0.4-10~K. $\lambda$ anomalies centered at 1.8(1), 2.5(1) and 1.8(1)~K were observed in the $C_{\textrm p}$ curves for NiBr$_2$(pyz)$_2$ (\textbf{2}), NiI$_2$(pyz)$_2$ (\textbf{3}) and Ni(NCS)$_2$(pyz)$_2$ (\textbf{4}), respectively. The lattice contributions ($C_\textrm{latt}$) to heat capacities are calculated by fitting the $C_{\textrm p}$ at high temperatures ($>10~\textrm{K}$) using a simple Debye fitting.\cite{MansonDaltTran2012} After subtracting the lattice contribution, the temperature dependence of magnetic entropy is calculated as shown in the inset to Fig.\,\ref{figCP1}, which exhibits the tendency to saturate to $R$ln(3) for all four compounds. This suggests that the $C_{\textrm{p}}$ anomaly stems from the $S = 1$ spin [Ni(II) ions] for \textbf{1}-\textbf{4}.

\begin{figure}[t]
\centering
\includegraphics[width=\columnwidth]{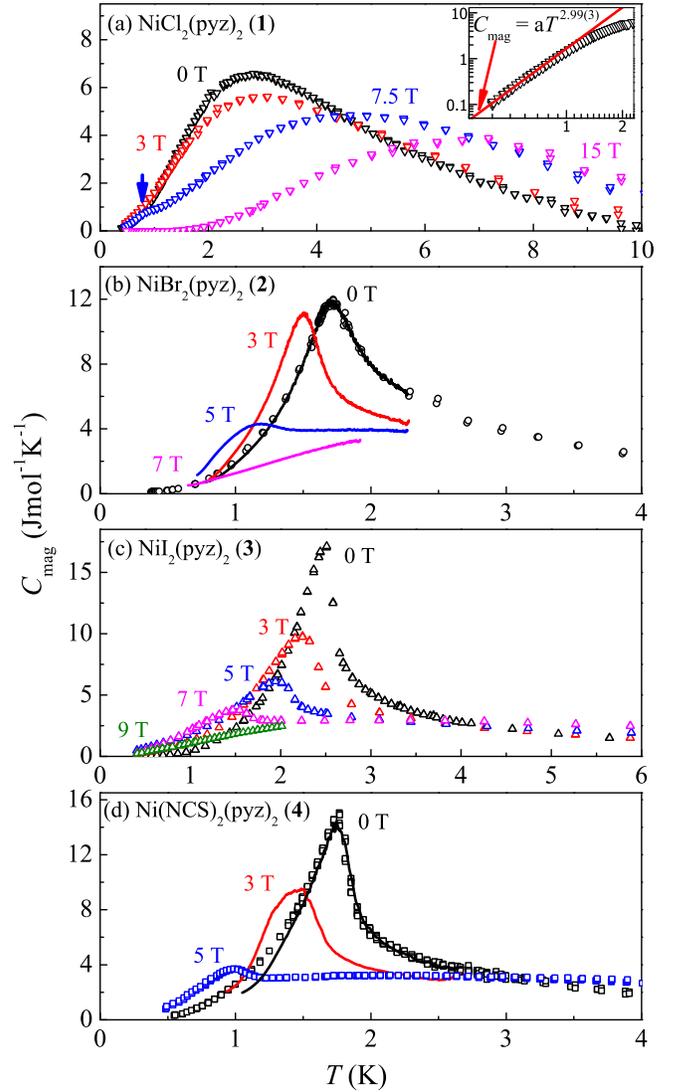}
\caption{$C_{\mathrm{mag}}$ versus $T$ for Ni$X_2$(pyz)$_2$ ($X$ = Cl (\textbf{1}), Br(\textbf{2}), I (\textbf{3}) and NCS (\textbf{4})) under various magnetic fields. The open symbols and solid curves corresponds to the data obtained by the traditional relaxation and dual-slope methods, respectively. Inset to (a): the low-temperature section of the zero-field $C_{\mathrm{mag}}$ for \textbf{1} plotted on a logarithmic scale. The red line is a fit to the spin-wave expansion, $C_{\mathrm{mag}} = aT^{d/n}$, for the $T < 0.6~\mathrm{K}$ data.}
\label{figCP2}
\end{figure}

The distinct $C_{\textrm{p}}$ anomalies for \textbf{2}-\textbf{4} are attributed to the antiferromagnetic (AFM) LRO of $S = 1$ spins. In low-dimensional antiferromagnets with strong spatial exchange anisotropy, $\lambda$ peaks are suppressed due to the onset of short-range ordering above $T_\textrm{N}$ which reduces the entropy change at the transition to LRO.\cite{SenguptaPRB2003} The presence of the $\lambda$ peaks indicates that \textbf{2}-\textbf{4} are close to 3D antiferromagnets in which the interactions in all directions, i.e. within and between the [Ni(pyz)$_2$]$^{2+}$ layers, are similar. On the other hand, the $C_\textrm{p}$ for NiCl$_2$(pyz)$_2$ (\textbf{1}) shows no sharp peak over the measured $T$-range. The broad $C_{\textrm{p}}$ peak in \textbf{1} can be explained by the thermal excitation among the $S = 1$ spin states (Schottky anomaly) and/or low dimensional spin correlations\cite{SenguptaPRB2003}. Unfortunately, we could not draw an unambiguous conclusion for the sign or the magnitude of $D$ for \textbf{1}. However, the hypothesized $D$ value (based on ESR and susceptibility measurements) is significantly stronger than the exchange interaction between Ni(II) ions (see below). Therefore, the thermal excitation among the $S = 1$ multiplet is expected to have marked contributions to the magnetic heat capacity of \textbf{1} at high temperatures. The magnetic contribution ($C_\textrm{mag}$) to the heat capacity for \textbf{1} is calculated by subtracting $C_\textrm{latt}$ from $C_\textrm{p}$ as shown in Fig.\,\ref{figCP2}(a). Below 0.6~K, $C_\textrm{mag}$ can be fitted to the spin-wave excitation, $C_\textrm{mag} \propto T^{d/n}$, with $d = 2.99(3)$ and $n = 1$ as shown in the inset to Fig.\,\ref{figCP2}(a). The $d$ value obtained from the low temperature fit is very close to the $T^3$ dependence expected for 3D AFM spin waves\cite{SoraiChemRev2006,HeatCapComment}. Hence, it is likely that \textbf{1} goes through a transition to LRO within the experimental temperature range. The lack of a $\lambda$-peak is indicative of the presence of significant spatial anisotropy in the magnetic interactions in \textbf{1}. Based on the comparison between \textbf{1} and \textbf{2} (\textbf{3}), we expect Q2D magnetism for \textbf{1} and $J_\textrm{pyz} \gg J_\perp$ (see more details in Sec.~\ref{discussion}), where $J_\textrm{pyz}$ is the intralayer interaction and $J_\perp$ is the interlayer interaction. For a layered Heisenberg $S = 1$ antiferromagnet, the $\lambda$-anomaly diminishes and becomes almost invisible when $J_\perp/J_\textrm{pyz} = 0.01$.\cite{JuhaszJungerPRB2009} In the case of \textbf{1}, the $J_\perp/J_\textrm{pyz}$ ratio at which the $\lambda$-anomaly vanishes is expected to deviate from 0.01 due to the presence of $D$ which may reduce the degrees of freedom of Ni(II) spins. Nevertheless, we expect $J_\perp$ to be at least an order of magnitude smaller than $J_\textrm{pyz}$ ($J_\perp/J_\textrm{pyz} < 0.1$) in order to account for the absence of a $\lambda$-anomaly in \textbf{1}. 

\begin{figure}[t]
\centering
\includegraphics[width=\columnwidth]{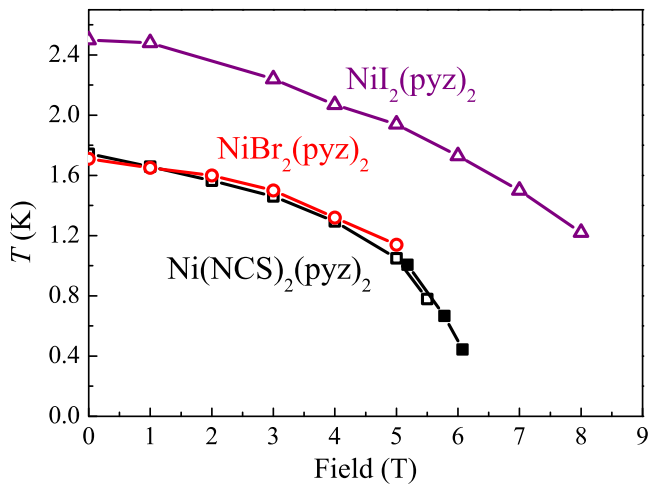}
\caption{Phase boundary for \textbf{2} ($\circ$), \textbf{3} ($\triangle$) and \textbf{4} ($\Box$ and $\blacksquare$) measured by heat capacity and MCE. The open symbols and the solid squares are extracted by heat capacity and MCE, respectively.}
\label{figCP3}
\end{figure}

Fig.~\ref{figCP2} shows the temperature dependence of $C_\mathrm{mag}$ at various magnetic fields. For \textbf{1}, a small shoulder develops below 2~K upon the application of a magnetic field up to 7.5~T (indicated by the arrow). Above 7.5~T, the broad peak for \textbf{1} moves to higher temperatures, which is due to the Zeeman splitting effect on the magnetic band structure. The field dependence of $C_{\textrm{p}}$ for \textbf{2}-\textbf{4} are similar to each other. The LRO temperature is suppressed by the application of magnetic fields. The phase diagrams for \textbf{2}-\textbf{4} are shown in Fig.\,\ref{figCP3}. The open symbols and solid squares are the phase boundary extracted by $C_\textrm{p}(T)$ and MCE, respectively.  The phase boundaries observed in \textbf{2} and \textbf{4} are commonly seen in the phase diagram of a 3D antiferromagnet. The amplitude of the specific-heat anomalies at zero field diminish from 17~kJ/mol (\textbf{3}) to 12~kJ/mol (\textbf{2}). In particular, \textbf{2} and \textbf{4} exhibit the same LRO temperature whereas the height of the $\lambda$-peaks is reduced from 15~kJ/mol (\textbf{4}) to 12~kJ/mol (\textbf{2}). The reduction in the amplitude of the $\lambda$-peak is often indicative of a reduction of the interlayer interaction.\cite{SenguptaPRB2003}


\subsection{Search for long range ordering with $\mu$SR}

\begin{figure}[b]
\begin{center}
\includegraphics[width=\columnwidth]{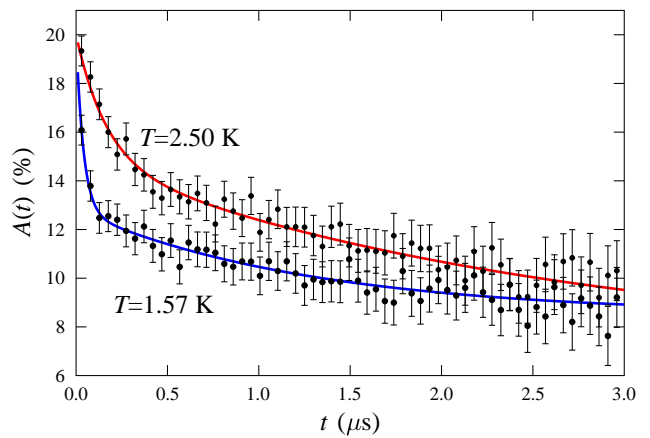}
\caption{Example ZF $\mu^{+}$SR data measured on NiBr$_{2}$(pyz)$_{2}$ ({\bf 2}) above and below the transition at 1.9(1)~K. The solid lines are fits of the data to Eq.~\ref{fiteq}.}
\label{figuSRspectra}
\end{center}
\end{figure}

Example $\mu$SR spectra measured on NiBr$_{2}$(pyz)$_{2}$  (\textbf{2}) are shown in Fig.\,\ref{figuSRspectra}. 
Across the measured temperature range $1.5 \leq T \leq 5$~K we observed monotonic relaxation with no resolvable oscillations in the spectra. (In fact we found that the form of the spectra for materials {\bf 1}-{\bf 3} all share the same form.) The spectra were found to be well described by the function
\begin{equation}
A(t) =A_{1}e^{-\lambda_{1} t} + A_{2}e^{-\lambda_{2} t} + A_{\mathrm{bg}},
\label{fiteq}
\end{equation}

\begin{figure*}
\begin{center}
\includegraphics[width=17cm]{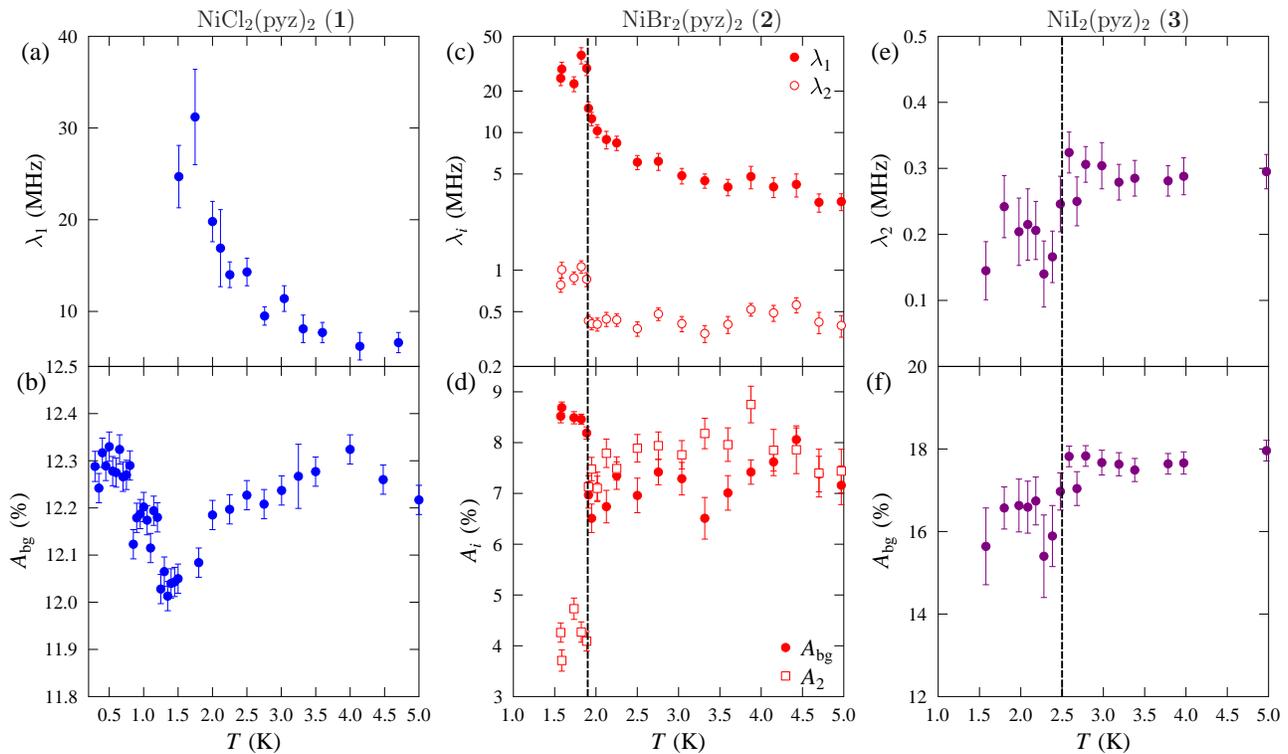}
\caption{The temperature evolution of selected parameters in Eq.~\ref{fiteq} for material {\bf 1} [(a) and (b)], {\bf 2} [(c) and (d)] and {\bf 3} [(e) and (f)]. Plot (b) shows that a broad minimum is observed in the non-relaxing component ($A_\mathrm{bg}$) for \textbf{1} around 1.5~K. Sharp discontinuities are observed in the fitted parameters for \textbf{2} and \textbf{3} [plots (c)-(f)] at 1.9~K and 2.5~K, respectively, indicating a magnetic transition at these temperatures. The vertical dash lines are guides for the eyes showing the temperatures at which magnetic ordering occurs in \textbf{2} and \textbf{3}.} 
\label{figuSRpara}
\end{center}
\end{figure*}

where the initial amplitude $A(0)$ was held fixed. $A_1$ and $A_2$ correspond to the fast and slow relaxing components, respectively. The temperature evolution of the fitted parameters for {\bf 2} is shown in Fig.~\ref{figuSRpara}(c) and (d). In both the spectra (Fig.~\ref{figuSRspectra}) and in the behavior of the fitted parameters [Fig.~\ref{figuSRpara}(c) and (d)] we see a sharp discontinuity on cooling through $T\approx 1.9$~K. This involves a decrease in the amplitude $A_{2}$ of the slowly relaxing component with relaxation rate $\lambda_{2}$, implying an increase in the amplitude $A_{1}$ of the component with relaxation rate $\lambda_{1}$. The fact that the non-relaxing component $A_{\mathrm{bg}}$ increases sharply implies a transition to a regime with a static distribution of local fields in the sample. This is because those muons whose spins lie parallel to the static local magnetic field at the muon site will not be relaxed\cite{SteelePRB2011} and will therefore contribute to the non-relaxing amplitude $A_{\mathrm{bg}}$. In addition, the relaxation rates would be expected to be proportional to  the second moment of the local magnetic field distribution $\langle B^{2} \rangle$. The rapid increase in relaxation rates $\lambda_{1}$ and $\lambda_{2}$ therefore probably implies an increase in the magnitude of the local magnetic fields at the muon sites. Taken together, these phenomena point towards a transition to a regime of 
magnetic order taking place at $T_{\mathrm{N}} = 1.9(1)$~K in {\bf 2}, which is in reasonable agreement with the peak in $C_{\mathrm{p}}$.

Measurements on NiI$_{2}$(pyz)$_{2}$ (\textbf{3}) were made using the MuSR spectrometer at ISIS. The pulsed muon beam at ISIS has a time width $\tau \approx 80$~ns, which limits the time resolution to below $\approx 1/\tau$. As a result, we are unable to  resolve the fast relaxation (with rate $\lambda_{1}$) that we considered in the data for material {\bf 2}, which manifests itself as missing asymmetry.  Instead we plot the slow relaxation rate [Fig.\,\ref{figuSRpara}(e)]
and the baseline asymmetry ($A_\mathrm{bg}$) [Fig.\,\ref{figuSRpara}(f)] which show discontinuities on magnetic ordering around a temperature $T_{\mathrm{N}}=2.5(1)$~K, in agreement with the anomaly in the heat capacity. 

Measurements were made on Ni(NCS)$_{2}$(pyz)$_{2}$ ({\bf 4}) using the ARGUS spectrometer at the ISIS facility. In this case the spectra showed weak exponential relaxation in the regime $0.35 \leq T \leq 4$~K with no discontinuities observed that would reflect the ordering temperature seen in the heat capacity at $T_{\mathrm{N}}=1.8$~K. It is unclear why the
muon should be insensitive to the ordering transition in this material, although we note the possibility of the muon forming bound states with the electronegative (NCS)$^{-}$ and therefore being insensitive to the ordering of the electronic moments. However, this was not the case in Fe(NCS)$_{2}$(pyz)$_{2}$ \cite{LancasterPhysicaB2006} where the spectra were of the same form as observed here for materials {\bf 1}-{\bf 3} and the magnetic ordering transition was observed. 

For measurements made on NiCl$_{2}$(pyz)$_{2}$ ({\bf 1}) using the GPS spectrometer, no sharp change in the form of the spectra is observed in the accessible temperature range $T>1.5$~K, although we saw a steep rise in the fast relaxation rate [Fig.~\ref{figuSRpara}(a)] as temperature is lowered below 2~K. In order to search for magnetic order in {\bf 1}, measurements were made down to 0.35~K using a sorption cryostat with the EMU spectrometer at ISIS. As in the case of material \textbf{1}, the ISIS resolution limit prevents us from resolving  fast relaxation in this case. Instead, it is instructive to follow $A_{\mathrm{bg}}$ as a function of temperature, shown in Fig.~\ref{figuSRpara}(b). On cooling we see a sharp decrease below 2~K, leading to a minimum in asymmetry centered around 1.5~K. The decrease in asymmetry on cooling is probably due to the increase in relaxation of the muon spins. This is followed by an increase at lower temperatures probably reflecting a regime where the moments are more static. It is possible that this minimum reflects a magnetic transition in material {\bf 1}, although the difference in the heat capacity for this compound compared to others in the series means that this is unlikely to be a transition to a regime of long-range magnetic order. Instead it is possible that the changes in the $\mu$SR spectra we observe in the 1.5--2~K region reflect a freezing-out of dynamic relaxation channels causing moments to become more static on the muon ($\mu$s) timescale.


\subsection{Electron spin resonance}


\begin{figure}[b]
\centering
\includegraphics[width=\columnwidth]{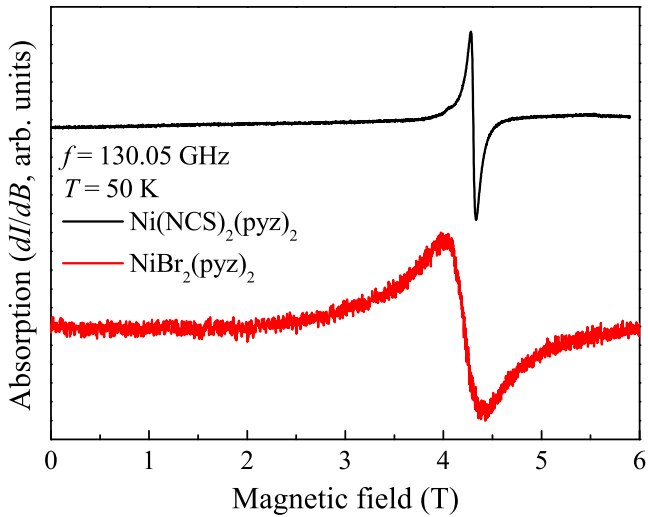}
\caption{Representative 130~GHz ESR spectra for \textbf{2} (red) and \textbf{4} (black) collected at 50~K. The absorption ESR  spectra are recorded in the first derivative mode.}
\label{figESR}
\end{figure}

Electron spin resonance (ESR) measurements were performed on powder samples of \textbf{1}-\textbf{4} to probe the ZFS and the $g$ factor associated with Ni(II) ions. A thorough search for ESR absorption in NiCl$_2$(pyz)$_2$ (\textbf{1}) at 130~GHz gave no indication for any ESR signal in the temperature range $1.9 \le T \le 300~{\textrm K}$, in contrast to \textbf{2}-\textbf{4}. The lack of ESR signal in \textbf{1} is indicative of the presence of a sizable ZFS ($|D| \ge 6.24~{\textrm K}$) for \textbf{1}. The representative ESR spectra for NiBr$_2$(pyz)$_2$ (\textbf{2}) and Ni(NCS)$_2$(pyz)$_2$ (\textbf{4}) at 50~K are shown in Fig.\,\ref{figESR}. The spectra were recorded in the first-derivative mode. A single ESR transition was observed for \textbf{2} and \textbf{4} up to 6~T. The broad ESR linewidth for \textbf{2} is likely due to structural-disorder-induced $g$-strain/$D$-strain\cite{AbragamEPRbook2012,ParkPRB2001} as shown by crystallography data. In the high temperature regime ($T \gg T_\textrm{N}$), the observed ESR signal corresponds to single-spin excitations associated with Ni(II). For $S = 1$ Ni(II) with a non-zero ZFS and/or anisotropic $g$ factor, a powder ESR spectrum is expected to show multiple transitions which correspond to the field being parallel/perpendicular to the magnetic-principle axis of Ni(II). The observation of a single transition in ESR spectra suggests that $D = 0$ as well as $g_x = g_y = g_z$ for Ni(II) ions in \textbf{2} and \textbf{4}. The center of the transition gives $g = 2.20(5)$ and $g = 2.16(1)$ for \textbf{2} and \textbf{4}, respectively. The ESR spectra for NiI$_2$(pyz)$_2$ (\textbf{3}) recorded at 130 GHz (not shown) only exhibit an extremely broad feature which is not applicable for a quantitative analysis.

\begin{figure}[t]
\centering
\includegraphics[width=\columnwidth]{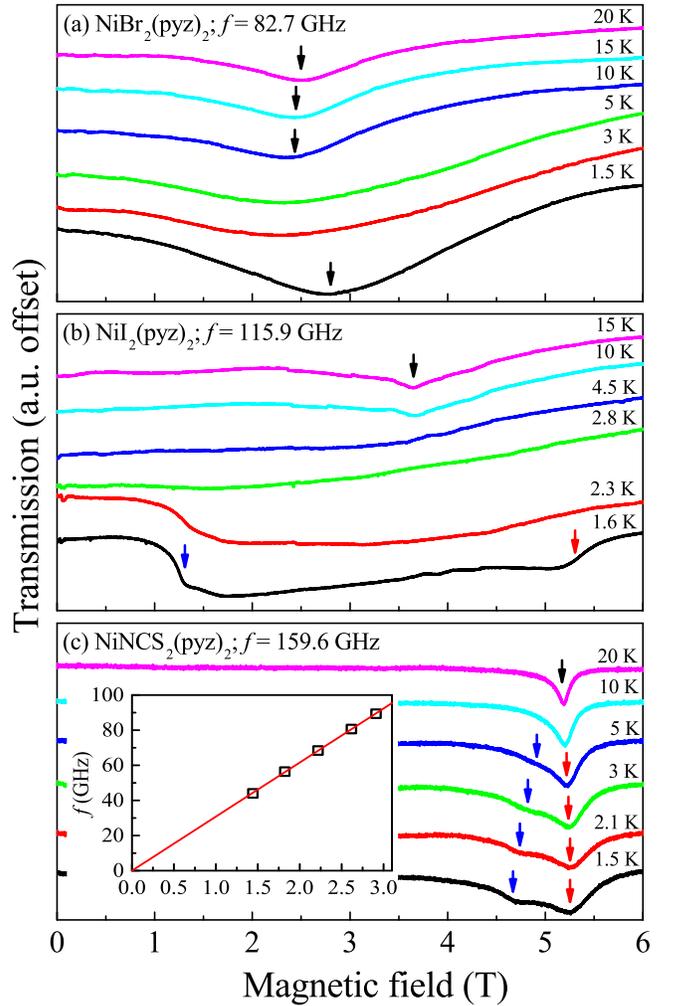}
\caption{Temperature dependence of the ESR spectra for powder samples of (a) NiBr$_2$(pyz)$_2$ (\textbf{2}), (b) NiI$_2$(pyz)$_2$ (\textbf{3}) and (c) Ni(NCS)$_2$(pyz)$_2$ (\textbf{4}) recorded at 82.7~GHz, 115.9~GHz and 159.6~GHz, respectively. The spectra are recorded in the transmission mode. The inset to (c) shows the frequency versus field plot for the ESR resonance observed in \textbf{4} at 50~K. The solid line correspond to a fit of the data with $g = 2.18(3)$ and $D = 0$.}
\label{figMVNA}
\end{figure}

Further variable frequency/temperature ESR measurements were performed on \textbf{2}-\textbf{4} in a broadband ESR spectrometer. Representative ESR spectra are shown in Fig.\,\ref{figMVNA}. The spectra were recorded in the transmission mode. The 20~K spectra for \textbf{2} and \textbf{4} [Fig.\,\ref{figMVNA}(a) and (c)] are consistent with the aforementioned 130~GHz results where a single transition was observed, suggesting $D = 0$ and $g_x = g_y = g_z$. Additional multi-frequency ESR measurements were performed on \textbf{4} to confirm the absence of $D$ in the compound [inset to Fig.\,\mbox{\ref{figMVNA}}(c)]. The 15~K spectrum for \textbf{3} [Fig.\,\ref{figMVNA}(b)] exhibits a broad feature which spreads over the entire field range (6~T).  This feature is reminiscent of a spectrum for $g = 2.27(8)$ and $D = 0$ Ni(II) ions. The broad linewidth associated with the ESR signal of \textbf{3} is likely due to $g$-strain/$D$-strain and/or the presence of non-Heisenberg interactions\cite{CastnerPRB1971} between Ni(II) ions (see below).

Upon cooling, the ESR resonance fields and linewidths for \textbf{2}-\textbf{4} show substantial variations as the temperature approaches the onset of LRO. The temperature dependence of the spectra above $T_\textrm{N}$ may be attributed to short-range spin correlations.\cite{NagataJPSJ1972,TazukeJPSJ1971} When the temperature approaches $T_\textrm{N}$, it is conceivable that small clusters of spins can be strongly correlated and exhibit properties that prefigure the long-range ordered behavior. At low temperatures, the spectra for \textbf{2}-\textbf{4} show distinct differences. For \textbf{2}, a single resonance was observed down to the base temperature. On the other hand, two resonances are observed in the low temperature spectra for \textbf{3} and \textbf{4}, as indicated by the blue and red arrows in Fig.\,\ref{figMVNA}(b) and (c). It is known that ESR probes antiferromagnetic resonances when $T < T_\textrm{N}$ where the multiple resonances corresponds to the applied field being parallel/perpendicular to the collective anisotropy field and/or different AFM modes in powder samples.\cite{KatsumataJPhysCondMatt2000} In either case, the observation of multiple ESR transitions in the low temperature spectra for \textbf{3} and \textbf{4} reveals the presence of a collective anisotropy field in these two compounds. Due to the fact that no single-ion ZFS was found for \textbf{3} and \textbf{4} at high temperatures, the collective anisotropy fields are likely due to non-Heisenberg interactions between Ni(II) ions. By contrast, the anisotropy field in \textbf{2} is likely to be negligible as only a single transition is observed down to the lowest temperature.

Quantitative calculations of the anisotropy fields in \textbf{3} and \textbf{4} are complicated by the fact that the transition temperatures are significantly affected by the applied field (see the phase diagram in Fig.\,\ref{figCP3}). In the experimental temperature regime, most low temperature spectra spread across the phase boundary which makes it very difficult to simulate the ESR spectra with any standard model. Qualitatively speaking, the spacing between the two resonances in \textbf{3} is almost four times of that of \textbf{4}, suggesting the presence of a stronger anisotropy field in \textbf{3} than \textbf{4}. This is confirmed by the spin-flop transition observed in the these two compounds (see below).
 

\subsection{Pulsed field magnetization}

\begin{figure}[b]
\centering
\includegraphics[width=\columnwidth]{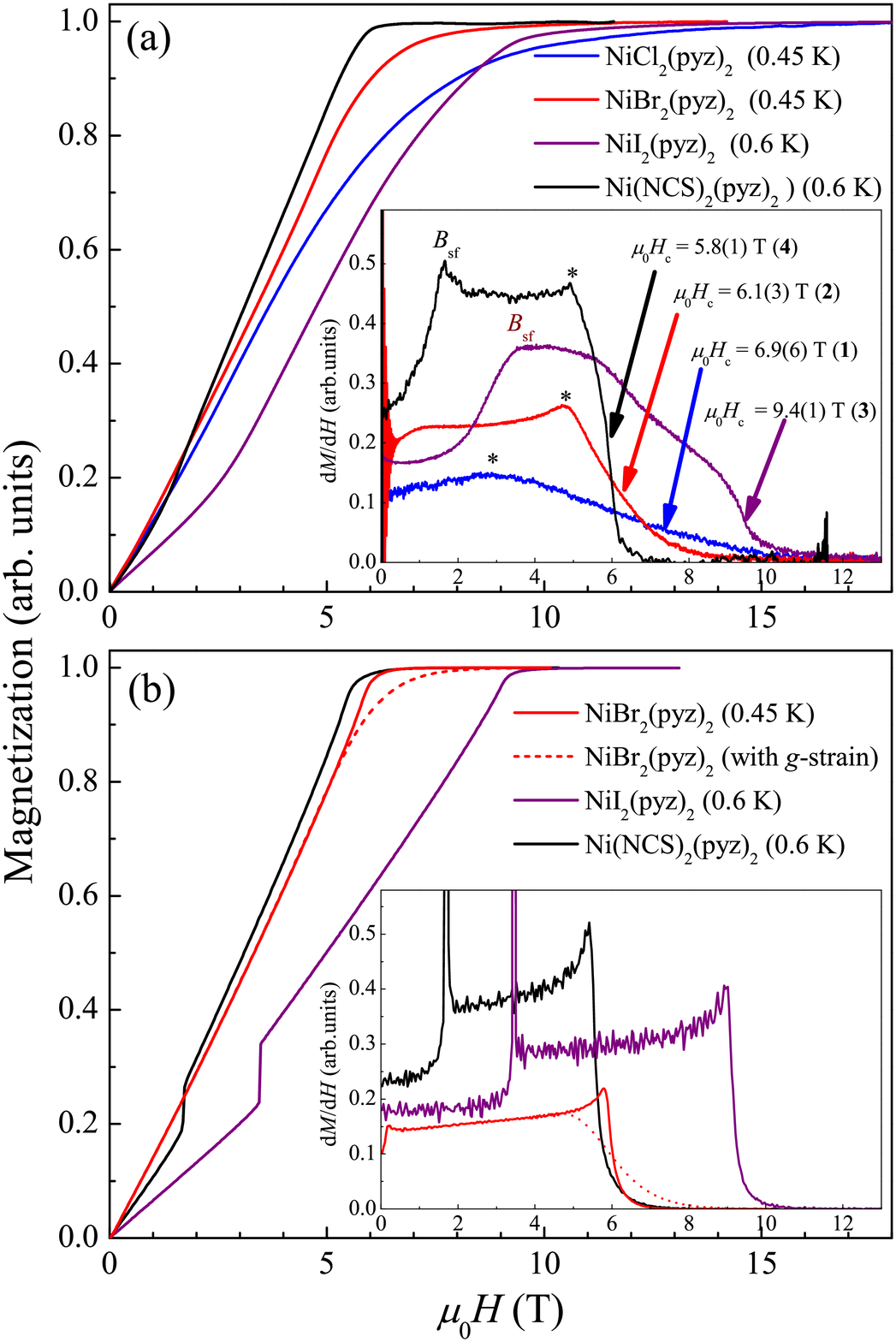}
\caption{(a) Main plot: Isothermal magnetization for NiCl$_2$(pyz)$_2$ (\textbf{1}), NiBr$_2$(pyz)$_2$ (\textbf{2}), NiI$_2$(pyz)$_2$ (\textbf{3}) and Ni(NCS)$_2$(pyz)$_2$ (\textbf{4}) acquired well below their ordering temperatures for \textbf{2}-\textbf{4}. Inset: $\mathrm{d}M/\mathrm{d}H$ plot showing the spin-flop transition ($B_\textrm{sf}$) and critical fields ($B_\textrm{c}$). (b) Main plot: Calculated magnetization $M$ for NiBr$_2$(pyz)$_2$ (red), NiI$_2$(pyz)$_2$ (purple) and Ni(NCS)$_2$(pyz)$_2$ (black) employing a $S = 1$ square lattice with interlayer interactions (Eq.\,\ref{QMC}). Inset: $\mathrm{d}M/\mathrm{d}H$ plot for the calculated magnetization.  The dash lines in both the main plot and the inset represent the simulation for NiBr$_2$(pyz)$_2$ including a broadening effect induced by $g$-strain.}
\label{figM}
\end{figure}

Magnetization versus field data ($M$ vs $H$) were recorded between 0.45~K and 10~K using pulsed-magnetic fields up to 60~T and are shown in Fig.\,\ref{figM}(a). At low temperatures, all compounds exhibit a slow initial rise in $M$ which gradually increases slope until the critical field ($H_{\textrm{c}}$) is approached. $\mu_0H_{\textrm{c}} = 6.9(6)$, 6.1(3) and 5.8(1)~T for NiCl$_2$(pyz)$_2$ (\textbf{1}), NiBr$_2$(pyz)$_2$ (\textbf{2}) and Ni(NCS)$_2$(pyz)$_2$ (\textbf{4}) respectively, as defined by the midpoint between the peak in $\mathrm{d}M/\mathrm{d}H$ (indicated by $\ast$ in the inset to Fig.\,\ref{figM}) and the region where $\mathrm{d}M/\mathrm{d}H$ remains essentially constant (inset to Fig.\,\ref{figM}). The slight concavity of the $M$ vs $H$ curve is expected for antiferromagnetic $S = 1$.\cite{WierschemJPhysConf2012} In the case of NiI$_2$(pyz)$_2$ (\textbf{3}), the $\mathrm{d}M/\mathrm{d}H$ curve exhibits extra steps between 6$\sim$10~T which may be attributed to non-Heisenberg exchange interactions as well as the polycrystalline nature of the sample. The presence of non-Heisenberg interactions can give rise to an anisotropic critical field, leading to extra steps at high fields in the $\mathrm{d}M/\mathrm{d}H$ curve of a powder sample. The critical field for \textbf{3} is defined by the midpoint between the last kink in $\mathrm{d}M/\mathrm{d}H$ and the region where $\mathrm{d}M/\mathrm{d}H$ drops to zero. It is noteworthy that due to the possibility of an anisotropic critical field, this assigned value (9.4(1)~T) for \textbf{3} may be an overestimation and actually correspond to the largest component of the anisotropic $H_\textrm{c}$.

For \textbf{3} and \textbf{4}, low-field anomalies occur at 3.46 and 1.68~T, respectively, which are attributed to a field induced spin-flop transition. It is well established\cite{AndersonPhysRev1964,JonghAdvPhys1974} that the spin-flop field $B_\mathrm{sf} = \mu_0H_\mathrm{sf}$ is related to the anisotropy field $H_\mathrm{A}$ and the exchange field $H_\mathrm{E}$ ($\approx H_\textrm{c}/2$) by $H_\textrm{sf}^2 = 2H_\textrm{E}H_\textrm{A} - H_\textrm{A}^2$. Based on this relation, the anisotropy fields are estimated to be 1.52 and 0.54~T for \textbf{3} and \textbf{4}, respectively. No evidence of a spin-flop transition was found for \textbf{1} and \textbf{2}. The magnetization data for \textbf{2}-\textbf{4} are consistent with the low-temperature ESR spectra, i.e., the anisotropy field of \textbf{2} is negligible whereas that of \textbf{3} is found to be significant. An intermediate anisotropy field is observed in \textbf{4}.

The rounded nature of $M$ in the vicinity of $H_{\textrm c}$ could be due to several reasons including the powdered nature of the samples, a sizable zero-field splitting and/or anisotropic $g$ factors. For \textbf{4}, the gradient of the $M(H)$ curve decreases rapidly until $M$ saturates at around 6~T. In comparison, the transitions from nearly linearly increasing to saturated behavior in the $M$ vs $H$ curves for \textbf{1} and \textbf{2} is broadened, as is often found in polycrystalline samples in several Ni(II)-based polymeric magnets. This difference in the transitions for \textbf{1}, \textbf{2} and \textbf{4} is in line with the ESR results. The ESR spectra for \textbf{4} are indicative of the absence of ZFS as well as an isotropic $g$ associated with Ni(II), leading to a sharp transition in the vicinity of $H_{\textrm c}$. Whereas in \textbf{1}, the lack of ESR signal up to 130~GHz ($= 6.24~\textrm{K}$) indicates the presence of a sizable ZFS in Ni(II) ($|D| \ge 6.24~{\textrm K}$), which leads to an extremely broad transition in the magnetization curve. For \textbf{2}, though $D = 0$ and $g$ is isotropic, the broad ESR linewidth implies a broad distribution of $g$ ($g$-strain), resulting in an intermediate broadened transition in its $M$ versus $H$ data.

Fig.\,\ref{figM}(b) shows the calculated magnetization for \textbf{2}-\textbf{4} at low $T$. The simulations are performed using the stochastic series expansion (SSE) method\cite{SyljuasenPRE2003} employing the following Hamiltonian:
\begin{multline}
\label{QMC}
{\cal H}=\sum_{\left<ij\right>_{xy}}J_\textrm{pyz}\left(S_{i}^{x}S_{j}^{x}+S_{i}^{y}S_{j}^{y}+\Delta S_{i}^{z}S_{j}^{z}\right)\\
+\sum_{\left<ij\right>_z}J_\perp\left(S_{i}^{x}S_{j}^{x}+S_{i}^{y}S_{j}^{y}+\Delta S_{i}^{z}S_{j}^{z}\right)-\sum_{i}{\vec B}\cdot\vec{S_{i}}.
\end{multline}
The simulations are performed with $J_\textrm{pyz} = 1.00~\textrm{K}$ and $J_\perp = 0.26~\textrm{K}$ for \textbf{2}, $J_\textrm{pyz} = 0.85~\textrm{K}$ and $J_\perp = 1.34~\textrm{K}$ for \textbf{3} and $J_\textrm{pyz} = 0.74~\textrm{K}$ and $J_\perp = 0.42~\textrm{K}$ for \textbf{4}. In the simulations, the ratio between $J_\textrm{pyz}$ and $J_\perp$ is fixed according to the magnetism dimensionality analysis (see Sec.~\ref{discussion} and Table~\ref{TableParameter}) while their values have been slightly fine tuned to match the experimental data. Additionally, we allowed an Ising-like interaction with $\Delta = 1.35$ and 1.20 for \textbf{3} and \textbf{4}, respectively, to account for the low field spin-flop transition. $\Delta = 1$ (Heisenberg interaction) for \textbf{2} as no collective anisotropy is observed. In the simulations, we obtained the powder averages by calculating the magnetization curves $M_x$ for ${\vec B}=B{\hat x}$ and $M_z$ for ${\vec B}=B{\hat z}$ then using the mean field relation $M_p=\frac{1}{3}M_z+\frac{2}{3}M_x$. In the calculation we neglected the demagnetizing field and assumed $B = \mu_0H$. 

As shown in Fig.\,\ref{figM}, a good agreement between the experiments and simulations is obtained for \textbf{2} and \textbf{4}. For \textbf{2}, the rounded feature of $M$ in the vicinity of $H_{\textrm c}$ can be reproduced by including a structural disorder induced $g$-strain which leads to a Gaussian distribution of the $g$ factor. The inclusion of the Ising-like interactions ($\Delta > 1$) leads to a spin-flop transition in \textbf{3} and \textbf{4}, as shown by the anomaly in $\mathrm{d}M/\mathrm{d}H$. However, the simulation for \textbf{3} does not show any obvious kink at high fields in $\mathrm{d}M/\mathrm{d}H$ with $\Delta$ alone. The Ising-like interactions in \textbf{3} give rise to a 0.2~T difference between the critical fields with $B\parallel z$ and $B\perp z$ which appears to be insufficient to explain the high-field feature in experiments, suggesting additional anisotropy terms are needed to explain the magnetization data for \textbf{3}. 

Further investigations are required to fully understand the spin-flop transition in \textbf{3}-\textbf{4}. The anisotropic part of the interaction, $J(\Delta-1)$, should be proportional to $(\Delta g/g)^2$,\cite{MoriyaPR1960} where $\Delta g$ is the $g$ anisotropy of Ni(II). Therefore, it seems to be contradictory to include an Ising-type interaction whereas no $g$-anisotropy was observed in the ESR data. We suspect that the single-ion anisotropy of Ni(II) is not fully resolved due to non-Heisenberg interactions which broaden the ESR spectra\cite{CastnerPRB1971}. Further experiments have been proposed on their magnetic diluted congeners, Zn$_{1-x}$Ni$_x$$X_2$(pyz)$_2$ ($x\ll 1$), for investigating the Ni(II) anisotropy.


\subsection{Magnetic susceptibility and density functional theory}

\begin{figure}[t]
\centering
\includegraphics[width=\columnwidth]{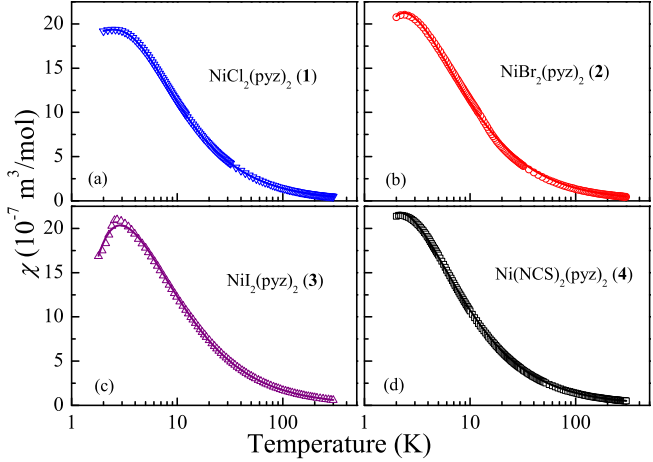}
\caption{Magnetic susceptibility data for powder sample of \textbf{1} (a), \textbf{2} (b), \textbf{3} (c) and \textbf{4} (d) collected with an applied magnetic field of 0.1~Am$^{-1}$. The solid lines represent fits of $\chi$ vs $T$ (see detailed discussion in the main text).}
\label{figChi}
\end{figure}

DC susceptibility measurements have been reported for \textbf{1} and \textbf{2} previously. The data were fitted to an anisotropic 2D model which gave $D = 7.92$ and 14.8~K, $zJ = 0.39$ and 0.95~K ($z=4$), $g=2.17$ and 2.31 for \textbf{1} and \textbf{2}, respectively.\cite{OtienoCanJChem1995,JamesAustJChem2002} Having discussed the magnetic dimensionality and the single-spin anisotropy from the aforementioned measurements, we now re-measure/analyze the DC susceptibility data for \textbf{1} and \textbf{2} [see Fig.~\ref{figChi}(a) and (b)]. Upon cooling from 300~K, $\chi(T)$ increases smoothly reaching a broad maximum near 2.6~K, 2.4~K, 2.7~K and 2.2~K for \textbf{1}, \textbf{2}, \textbf{3} and \textbf{4}, respectively, and then drops slightly as the temperature is lowered to 2~K. This behavior can be caused by concomitant antiferromagnetic (AFM) coupling between $S = 1$ Ni(II) sites and/or ZFS of the spin ground state. Curie-Weiss fits of the reciprocal susceptibility in the temperature range of $50<T<300~\textrm{K}$ lead to $g = 2.17(7)$ and $\theta = -3.51(23)~\textrm{K}$ (\textbf{1}), $g = 2.10(9)$ and $\theta = -3.20(36)~\textrm{K}$ (\textbf{2}), $g = 2.41(3)$ and $\theta = -5.02(6)~\textrm{K}$ (\textbf{3}) and $g = 2.10(4)$ and $\theta = -4.00(23)~\textrm{K}$ (\textbf{4}). In the absence of single-ion anisotropy, the negative Curie-Weiss temperatures would indicate the presence of AFM interactions in \textbf{1}-\textbf{4}. The fitted $g$ values for \textbf{2} and \textbf{4} are in good agreement with the ESR results. The fitted $g$ value for \textbf{3} deviates from the ESR result ($g = 2.27$) and appears to be too large for Ni(II). It is well known that the $g$ factor obtained from susceptibility can be affected by many experimental parameters, e.g. errors in the sample mass, whereas ESR gives a direct measurement for the $g$ factor. Therefore, for \textbf{2}-\textbf{4}, the $g$ factors extrapolated from the ESR data were used in the following data analysis. 

Based on the information obtained from the heat capacity and ESR studies, the $\chi(T)$ data for \textbf{2}-\textbf{4} were fitted to an $S = 1$ simple cubic Heisenberg model, $\hat{H} = J\sum_{\langle i,j \rangle} {{\hat{S}_i\cdot\hat{S}_j}}$. This model assumes that (a) the intra/interlayer interactions are the same ($=J$) and (b) the number of nearest magnetic neighbors, $z$, is 6, both of which may be oversimplifications. As we will mention in the discussion section, this model cannot account for the ordering temperature. Nevertheless, we can still use it to compare $zJ$ with the pulsed field magnetization data. Fig.\,\ref{figChi} shows the data and fits for \textbf{2}-\textbf{4} over the entire temperature range with the fitting parameters of $J = 0.82(5)~\textrm{K}$ (\textbf{2}), $J = 1.00(4)~\textrm{K}$ (\textbf{3}) and $J = 0.75(2)~\textrm{K}$ (\textbf{4}). These interactions would predict critical fields of $\mu_0H_\textrm{c}=6.66$, 8.4 and 6.2~T for \textbf{2}, \textbf{3} and \textbf{4} ($g = 2.20$ (\textbf{2}), $g=2.27$ (\textbf{3}) and 2.16 (\textbf{4}) from the ESR data), respectively. The estimated critical fields for \textbf{2} and \textbf{4} are in excellent agreement with the pulsed field data. The estimated critical field for \textbf{3} is slightly less than that measured in the magnetization data. However, as we mentioned in the previous section, the possibility of an anisotropic $H_\textrm{c}$ may lead to an overestimation of that in the magnetization data, which could account for this difference.

The susceptibility for \textbf{1} was fitted employing an anisotropic 2D model [Fig.\,\ref{figChi}(a)].\cite{CarlinChemRev1986} The fit gives $zJ_{\textrm{pyz}} = 1.97(4)~\textrm{K}$, $D = 8.03(16)~\textrm{K}$ and $g=2.15(5)$. Taking $z=4$ (for Q2D model), $J_\textrm{pyz} = 0.49(1)~\textrm{K}$ which is almost a half of that in \textbf{2}-\textbf{4}. The fitted easy plane type anisotropy $D=8.03~\textrm{K}$ gives rise to a broad peak (Schottky anomaly) around 3~K which coincides with the broad feature in $C_\textrm{p}$ for \textbf{1}. However, extrapolating $D$ and $J$ simultaneously from powder magnetic data can often be unreliable. The result is not unique and varies dramatically depending on the model employed in the analysis. In fact, it is possible to obtain a reasonable fit with the simple cubic 3D Heisenberg model with $J = 0.91(3)~\textrm{K}$. Because single crystals for \textbf{1} are currently unavailable, it is not possible to distinguish between the parallel and perpendicular susceptibilities in order to uniquely determine the sign and magnitude of $D$.

As an additional evaluation of the magnetic interactions, density functional theory (DFT) calculations were performed using the room temperature structural data for \textbf{1}-\textbf{4}. The magnetic interactions through the pyz bridges are modeled by the dinuclear fragments, (pyz)$_3$Ni$X_2$($\mu$-pyz)Ni$X_2$(pyz)$_3$, consisting of two (pyz)$_3$Ni$X_2$ segments connected by a bridging pyz ligand ($\mu$-pyz), which mediates the intralayer interaction $J_\mathrm{pyz}$. The calculations give weak AFM interactions mediated by Ni--pyz--Ni bonds throughout all compounds as expected. $J_\mathrm{pyz}$ are calculated to be 1.85, 2.41 and 3.16~K for compounds \textbf{1}, \textbf{2} and \textbf{3}, respectively. Separate DFT calculations were performed for \textbf{4} due to its lower symmetry ($C2/m$ vs. $I4/mmm$ for \textbf{1}-\textbf{3}). In general, the adjoining orthogonal pyz bridges in \textbf{4} afford different magnetic interactions depending on whether the Ni--Ni linkage lies in or perpendicular to the Ni--NCS planes. Therefore, DFT calculations for \textbf{4} were performed with both configurations to investigate the influence of the NCS ligand orientation onto $J_\mathrm{pyz}$. A small difference in $J_\mathrm{pyz}$ was found for these two configurations with $J_\mathrm{pyz}$ calculated to be 1.65 and 1.71~K for the Ni--Ni axis in and perpendicular to the Ni-NCS planes, respectively. The calculation for \textbf{4} suggests that $J_\mathrm{pyz}$ is almost independent of the orientation of the NCS ligands; hence, it is reasonable to treat the [Ni(pyz)$_2$]$^{2+}$ layers in \textbf{4} as magnetic square lattices in the data analysis.


\section{Discussion}
\label{discussion}

All of the four compounds share similar extended polymeric structures consisting of 2D square [Ni(pyz)$_2$]$^{2+}$ sheets in the $ab$-plane with the $X$ ligands acting as spacers between layers. The Ni-Ni separations are similar along the Ni-(pyz)-Ni bridges. There is little variation of the closest interlayer Ni-Ni distance across all four compounds (7.32~\AA\ for NiCl$_2$(pyz)$_2$ ({\textbf 1}), 7.54~\AA\ for NiBr$_2$(pyz)$_2$ ({\textbf 2}), 7.90~\AA\ for NiI$_2$(pyz)$_2$ (\textbf{3}) and 7.23~\AA\ for Ni(NCS)$_2$(pyz)$_2$ ({\textbf 4})). The difference in the magnetism of \textbf{1}-\textbf{4} clearly highlights the selection of the $X$ ligand can lead to significant changes in both the single-ion anisotropy and the magnetic dimensionality in this Ni$X_2$(pyz)$_2$ family. 

\begin{table*}[t]
\caption{The compounds studied in this work. The $J_\mathrm{pyz}$, $D$ and $g$ for NiCl$_2$(pyz)$_2$ (\textbf{1}) are obtained by fitting the DC susceptibility to an anisotropic 2D model while its $J_\perp$ is estimated based on the heat capacity data (see Sec.~\ref{SectionCp}). The $g$ values obtained via the ESR data and fitting the susceptibility are both listed in the table for comparison. The parameters for NiBr$_2$(pyz)$_2$ (\textbf{2}), NiI$_2$(pyz)$_2$ (\textbf{3}), Ni(NCS)$_2$(pyz)$_2$ (\textbf{4}) are determined by the analysis based on the heat capacity, ESR and pulsed magnetic field data (see Sec.\,\ref{discussion}).}
\centering
\begin{tabular}{l c c c c c c c}
\hline
 & \quad $J_\textrm{pyz}$~(K) & $J_\perp$~(K) & $D$~(K) & $g$ ($\chi(T)$) & $g$ (ESR) \quad & $T_\mathrm{N}$ (K) & $\mu_0H_\mathrm{c}$ (T)\\
\hline
NiCl$_2$(pyz)$_2$ (\textbf{1}) & \quad $0.49 \pm 0.01$ & $<0.05$ & $8.03\pm 0.16$ & $2.15\pm 0.05$ & n/a & n/a & $6.9\pm 0.6$\\
NiBr$_2$(pyz)$_2$ (\textbf{2})& \quad $1.00\pm 0.05$ & \quad $0.26\pm 0.05$ & 0 & $2.10\pm 0.09$ & $2.20\pm 0.05$ & $1.8\pm 0.1$ & $6.1\pm 0.3$\\
NiI$_2$(pyz)$_2$ (\textbf{3})&  $<1.19$ &  $>1.19$ & 0 & $2.41\pm 0.03$ & $2.27\pm 0.08$ & $2.5\pm 0.1$ & $9.4\pm 0.1$\\
Ni(NCS)$_2$(pyz)$_2$ (\textbf{4})& \quad $0.82\pm 0.05$ & \quad $0.47\pm0.05$ & 0 & $2.10\pm 0.04$ & $2.16\pm 0.01$ & $1.8\pm 0.1$ & $5.8\pm 0.1$\\
\hline
\end{tabular}
\label{TableParameter}
\end{table*}

Thorough investigations have been performed to quantify the magnetic interaction through $X$-bridges in Cu$X_2$(pyz) compounds ($X$ = F, Cl, Br and NCS).\cite{SchlueterInorgChem2012,LapidusChemComm2013,BordalloPolyhedron2003,LancasterJPhysCondMatt2004} The Cu$X_2$(pyz) compounds possess 2D rectangular lattices which are characterized by Cu-pyz-Cu chains linked by Cu-$X_2$-Cu bridges. We briefly review the interactions through the Cu-$X_2$-Cu bridges since they are likely related to the interlayer interactions through the $X$ ligands in compounds \textbf{1}-\textbf{4}. In Cu$X_2$(pyz) compounds, the AFM interactions through Cu-$X_2$-Cu bonds were found in the descending order of magnitude: Br$>$Cl$>$F$>$NCS. In particular, Cu(NCS)$_2$(pyz) presents itself as a nearly ideal Q1D AFM chain with the 1D interactions mediated through the Cu-pyz-Cu bridges. $\mu$SR measurements for Cu(NCS)$_2$(pyz) show no evidence for LRO above 0.35~K which is indicative of extremely weak interchain-interactions ($< 0.13~\textrm{K}$) through the Cu-(NCS)$_2$-Cu bonds.\cite{LancasterJPhysCondMatt2004} Therefore, it is at first sight surprising to see that  Ni(NCS)$_2$(pyz)$_2$ ({\textbf 4}) shows a strong $\lambda$ anomaly as the interlayer interactions via the NCS$^-$ ligands are expected to be small. On the other hand, the difference between  NiCl$_2$(pyz)$_2$ ({\textbf 1}) and NiBr$_2$(pyz)$_2$ ({\textbf 2}) may be explained by the previous studies with the less efficient Cl pathways leading to Q2D magnetism in \textbf{1}. The results for NiI$_2$(pyz)$_2$ (\textbf{3}) are in line with this hypothesis that the larger I$^-$ ions can form more efficient exchange pathways between [Ni(pyz)$_2$]$^{2+}$ layers, leading to stronger interlayer interactions. Consequently, a larger $\lambda$-anomaly and a higher $B_\textrm{c}$ are observed in the $C_\textrm{p}$ and the magnetization data.

A similar $\lambda$-anomaly in $C_{\textrm{p}}$ was observed in a compound isomorphous to \textbf{4}, Fe(NCS)$_2$(pyz)$_2$, which is regarded as an Ising Q2D antiferromagnet.\cite{BordalloPRB2004} In Fe(NCS)$_2$(pyz)$_2$, although long-range order is achieved below 6.8~K, its critical parameters are ideally close to those expected for Q2D Ising systems. In the case of \textbf{2}, the scenario for an Ising Q2D antiferromagnet is excluded due to the facts that (a) the ZFS of the Ni(II) ions in \textbf{2} are found to be negligible and (b) both the ESR and magnetization data show no evidence of a collective anisotropic field at low temperatures. For \textbf{3} and \textbf{4}, the absence of single-ion anisotropy in their paramagnetic phase is also unfavorable of extreme Ising Q2D antiferromagnets. In particular, the phase boundary of \textbf{4} is similar to that of 3D antiferromagnets, providing additional support for 3D antiferromagnetism in \textbf{4}. Therefore, it is most likely that the $X^-$ ligands serve as bridging ligands in \textbf{2}-\textbf{4} which mediate interlayer interactions that are comparable to the intralayer interactions, leading to AFM long range order. The difference between the NCS$^-$ bridges in Cu(NCS)$_2$(pyz) and \textbf{4} remain to be examined. The shortest Ni-S distance in \textbf{4} is 4.719~\AA\, which is unlikely to form a direct Ni-S exchange pathway. Therefore, the interlayer interactions in \textbf{4} are likely to be mediated through electron density overlapping between NCS$^-$ ligands connected to Ni(II) ions in adjacent layers. 

In discussing the susceptibility for \textbf{1}-\textbf{4}, a simple cubic model was employed for the data analysis. However, the legitimacy of using such a model needs to be justified. It is clear that each Ni(II) ion has four magnetic neighbors in its [Ni(pyz)$_2$]$^{2+}$ plane for all four compounds. However, it is not straightforward to tell the number of magnetic neighbors in the adjacent planes from the crystal structures. For \textbf{1}-\textbf{3}, each Ni(II) ion has 8 equally spaced neighbors in the adjacent planes. In the case of a perfect tetragonal space group, this gives 8 equivalent magnetic neighbors in the adjacent planes for a Ni(II) site, leading to frustration of the minimum-energy configuration if the interactions within the [Ni(pyz)$_2$]$^{2+}$ planes are antiferromagnetic.\cite{RastelliJPhysCondMatt1990} In which case, \textbf{1}-\textbf{3} would only show two-dimensional order within the [Ni(pyz)$_2$]$^{2+}$ planes and the $\lambda$-anomaly would be significantly suppressed, contrary to the experimental observations. Therefore, we speculate the frustration is relieved via breaking of the tetragonal symmetry, possibly due to the structural disorder of the pyz rings, resulting in 3D LRO in \textbf{2} and \textbf{3}. The breaking of the tetragonal symmetry should give rise to four inequivalent interlayer interactions in \textbf{1}-\textbf{3} with one of them being stronger than the others. \textbf{4} crystallizes in a monoclinic space group where one would expect four inequivalent interlayer interactions based on its structure. Therefore, it is reasonable to assume that the interlayer interactions are dominated by one particular pathway in \textbf{1}-\textbf{4} and each Ni(II) ion has two magnetic neighbors in the adjacent planes (one in the plane above/below). Although this is probably an oversimplification, it is the simplest model one can adopt and is consistent with the experimental results.

The critical fields measured in the pulsed magnetic field data provide a reliable way for probing the interactions between Ni(II). Here we focus on \textbf{2}-\textbf{4} in which no single-ion ZFS was observed in ESR. Consequently, $B_\textrm{c} = \mu_0H_\textrm{c}$ solely depends on the intra- and interlayer interactions. The critical field for \textbf{1} depends on both $D$ and $J$ and it is not possible to deconvolute them from pulsed field data alone. For quantitative calculations of the intra-/inter-layer interactions, the critical fields and the N\'eel temperatures for \textbf{2}-\textbf{4} are analyzed with a Q2D Heisenberg model. For $S = 1$ Q2D Heisenberg antiferromagnets, the critical field is
\begin{equation}
\label{Q2DM}
\mu_\textrm{B}gB_\textrm{c} = 8J_\textrm{pyz}+4J_\perp,
\end{equation}
where $J_\perp$ is the interlayer interaction. Yasuda \textit{et al} proposed an empirical correlation\cite{YasudaPRL2005} between the $T_\textrm{N}$ and the interactions based on Quantum Monte Carlo calculations for $S =1$ Q2D Heisenberg antiferromagnets:
\begin{equation}
\label{TnQ2D}
T_\textrm{N} = 4\pi\times 0.68J_\textrm{pyz}/[3.12-\textrm{ln}(J_\perp/J_\textrm{pyz})].
\end{equation}
Eq.\,\ref{TnQ2D} is valid in the range $0.001 \le J_\perp/J_\textrm{pyz} \le 1$. In the analysis we assumed $\Delta = 1$ due to the lack of theoretical study for the correlation between $\Delta$ and $T_\mathrm{N}$ in $S = 1$ antiferromagnets. Applying Eq.\,\ref{Q2DM} and Eq.\,\ref{TnQ2D} to \textbf{2}-\textbf{4}, it is found that the experimental results for \textbf{2} and \textbf{4} can be accounted for with the following parameter sets: $J_\textrm{pyz} = 1.0$ and $J_\perp = 0.26~\textrm{K}$ for \textbf{2} and $J_\textrm{pyz} = 0.82$ and $J_\perp = 0.47~\textrm{K}$ for \textbf{4}. The obtained $J_\textrm{pyz}$'s are similar for \textbf{2} and \textbf{4}, which is consistent with the structural similarities between their [Ni(pyz)$_2$]$^{2+}$ layers. $J_\perp/J_\textrm{pyz} = 0.26$ and 0.57 for \textbf{2} and \textbf{4}, respectively, indicating \textbf{2} is a 3D antiferromagnet which prefigures some Q2D magnetism whereas \textbf{4} is more similar to an ideal 3D antiferromagnet in which the intra- and inter-layer interactions are identical. The difference in  $J_\perp/J_\textrm{pyz}$ explains the reduction of the $\lambda$-anomaly in \textbf{2}. On the other hand, no $J_\perp$ and $J_\textrm{pyz}$ can satisfy Eq.\,\ref{Q2DM} and Eq.\,\ref{TnQ2D} simultaneously for \textbf{3}, suggesting it does not fall into the category of Q2D antiferromagnet. We suspect that the large I$^-$ ligands form efficient exchange pathways which propagate strong interlayer interactions, leading to $J_\perp > J_\textrm{pyz}$  in \textbf{3}. Hence, its LRO temperature and critical field cannot be interpreted as a Q2D antiferromagnet. Due to lack of theoretical study for $S = 1$ antiferromagnet with $J_\perp > J_\textrm{pyz}$, it is difficult to calculate $J_\perp$ and $J_\textrm{pyz}$ separately. In the case of an ideal 3D antiferromagnet,  $J_\perp = J_\textrm{pyz} = 1.19~\textrm{K}$ for \textbf{3}. With $J_\perp > J_\textrm{pyz}$, Eq.\,\ref{Q2DM} suggests $J_\textrm{pyz} < 1.19~\textrm{K}$ for \textbf{3}. However, among all the four compounds, \textbf{3} exhibits the strongest $\lambda$-anomaly, indicating it is expected to be reasonably close to a 3D antiferromagnet. Accordingly, we expect $J_\textrm{pyz}$ for \textbf{3} should be in the vicinity of 1~K. The parameters for \textbf{1}-\textbf{4} are summarized in Table~\ref{TableParameter}.

Finally, we compare the results for \textbf{1}-\textbf{4} with [Ni(HF$_2$)(pyz)$_2$]$Z$ ($Z$ = PF$_6^-$ and SbF$_6^-$). The 2D [Ni(pyz)$_2$]$^{2+}$ layers found in \textbf{1}-\textbf{4} exhibit very similar geometrical parameters to those of [Ni(HF$_2$)(pyz)$_2$]$Z$. The [Ni(HF$_2$)(pyz)$_2$]$Z$ compounds were found to be quasi-1D magnets composed of Ni-FHF-Ni chains ($J_{\textrm{1D}}$) with inter-chain coupling ($J_{\perp}$) mediated by Ni-pyz-Ni linkages. The interaction parameters were not determined due to difficulties in distinguishing between  $J_{\textrm{1D}}$, $J_{\perp}$ and $D$ from pulsed field data as above. The couplings through Ni-pyz-Ni bridges in \textbf{2}-\textbf{4} are found in the vicinity of 1~K, which are significantly smaller compared with $J_{\textrm{1D}}$ obtained in [Ni(HF$_2$)(pyz)$_2$]$Z$. Such results are consistent with the Q1D magnetism of [Ni(HF$_2$)(pyz)$_2$]$Z$. Our study also shows that the selection of the axial $X^-$ ligands can substantially vary the ZFS of Ni(II) as well as introduce non-Heisenberg interactions between Ni(II) ions, leading to different magnetic ground state structures in Ni(II) based magnets.

\section{Summary}

Four Ni(II) based coordination polymers are prepared and their structures are carefully examined. NiCl$_2$(pyz)$_2$ (\textbf{1}), NiBr$_2$(pyz)$_2$ (\textbf{2}), NiI$_2$(pyz)$_2$ (\textbf{3}) and Ni(NCS)$_2$(pyz)$_2$ (\textbf{4}) feature 2D square [Ni(pyz)$_2$]$^{2+}$ planes stacking along the $c$-axis spaced by $X$-ligands ($X = $Cl, Br, I or NCS). The heat capacity measurements are indicative of the presence of long-range order for \textbf{2}-\textbf{4} as well as Q2D magnetism for \textbf{1}. The $\mu$SR data for \textbf{1} suggest there seems to be a transition occurs at 1.5~K. The single-ion magnetic properties of \textbf{2}-\textbf{4} are measured by ESR where no evidence of ZFS was found. The pulsed-field magnetization data show the critical fields for \textbf{1}-\textbf{4} vary from 5.8~T to 9.4~T which are significantly smaller than those for [Ni(HF$_2$)(pyz)$_2$]$Z$ ($Z$ = PF$_6^-$ and SbF$_6^-$). Taken together, the magnetic property measurements reveal the interlayer interaction can be suppressed by the choice of the $X$ ligand. Despite the differences in the interlayer interactions, the Ni-pyz-Ni interactions in \textbf{2}-\textbf{4} remain largely unaltered and are found to be in the vicinity of 1~K. This result is in keeping with the prominent $\lambda$-anomaly in the heat capacity data and an excellent agreement for $T_\textrm{N}$ is obtained between experimental results and QMC predictions for \textbf{2} and \textbf{4}. The obtained $J_\textrm{pyz}$ values are consistent with the Q1D magnetism found in the [Ni(HF$_2$)(pyz)$_2$]$Z$ family. \textbf{1} possesses a finite ZFS and reduced magnetic dimensionality. This study, in combining with the previous works for the [Ni(HF$_2$)(pyz)$_2$]$Z$ family, reveals that prudent ligand choice may allow for systematically tuning the interlayer interaction between [Ni(pyz)$_2$]$^{2+}$ planes, permitting the preselection of Q1D, Q2D and 3D magnetism.

In addition to controlling the magnetic dimensionality, lattice randomness in low-dimensional $S = 1$ antiferromangets can lead to a highly nontrivial phase diagram.\mbox{\cite{RoscildePRL2007}} Such randomness can be introduced in molecule-based magnets by doping the system with diamagnetic ions, e.g. Zn(II), and the concentration of dopants can be controlled in the synthesis. The compounds studied in this work offer a promising opportunity for studying the effect of lattice randomness and other cooperative phenomena.

Improvements in the experimental testing of low-dimensional $S = 1$ antiferromagnets require better models for understanding the underlying physics. Specifically, a model for calculating the ordering temperature considering both the influence of the ZFS parameter \textit{D} and the exchange anisotropy is strongly desired for interpreting the experimental data. In addition, further DFT studies are required for a better appreciation of the mechanisms of the magnetic interactions as well as the ZFS of Ni(II). Such studies allow the prediction of the magnetic properties based on the crystalline structures, which can be anticipated with a high level of predictability in magnetic crystal engineering, and, therefore, raise the possibility of generating molecule-based magnets for better tests of the theories of low-dimensional magnetism.

\section{Acknowledgment}

A portion of this work was performed at the National High Magnetic Field Laboratory, which is supported by National Science Foundation Cooperative Agreement No. DMR--1157490, the State of Florida, and the U.S. Department of Energy (DoE) and through the DoE Basic Energy Science Field Work Proposal ``Science in 100 T''.  Work at EWU was supported by the National Science Foundation under grant no. DMR-1306158. Work in the UK is supported by the EPSRC and JS thanks Oxford University for the provision of a Visiting Professorship. Part of this work was carried out at the Swiss Muon Source, Paul Scherrer Institut, CH and at the ISIS Facility, STFC Rutherford Appleton Laboratory, UK. We are grateful to Alex Amato for technical assistance. Use of the Advanced Photon Source at Argonne National Laboratory was supported by the U. S. Department of Energy (DoE), Office of Science, Office of Basic Energy Sciences, under Contract No. DE-AC02-06CH11357. Use of the National Synchrotron Light Source, Brookhaven National Laboratory, was supported by the U.S. DoE, Office of Basic Energy Sciences, under Contract No. DE-AC02-98CH10886.

\bibliography{NiX2ref}

\begin{thebibliography}{81}
\bibitem [{Hea()}]{HeatCapComment}%
  \BibitemOpen
  \href@noop {} {}\bibinfo {note} {The lattice contribution to the heat
  capacity calculated by the Debye model is proportional to $T^3$, which has a
  power-law dependence similar to that of the low-temperature $C_\mathrm{mag}$.
  However, they are characterized by very different energy scales and at low
  temperatures ($T < 2~\mathrm{K}$) $C_\mathrm{mag} \gg
  C_\mathrm{latt}$.}\BibitemShut {Stop}%
\end{thebibliography}

\end{document}